\documentclass[11pt]{article}
\usepackage{amsmath,amssymb}
\newcommand{\ft}[2]{{\textstyle\frac{#1}{#2}}}

\newsavebox{\uuunit}
\sbox{\uuunit}
    {\setlength{\unitlength}{0.825em}
     \begin{picture}(0.6,0.7)
        \thinlines
        \put(0,0){\line(1,0){0.5}}
        \put(0.15,0){\line(0,1){0.7}}
        \put(0.35,0){\line(0,1){0.8}}
       \multiput(0.3,0.8)(-0.04,-0.02){10}{\rule{0.5pt}{0.5pt}}
     \end {picture}}

\allowdisplaybreaks

\numberwithin{equation}{section}

\begin{document}
\begin{titlepage}
\begin{center}
\hfill ITP-UU-10/10 \\
\hfill Nikhef-2010-006 \\
\vskip 6mm

{\Large \textbf{BPS black holes, the Hesse potential,\\ and the
    topological string }}
\vskip 8mm

\textbf{G.L.~Cardoso$^{a}$, B.~de~Wit$^{b,c}$ and S.~Mahapatra$^{d}$}

\vskip 4mm
$^a${\em  
CAMGSD, Departamento de Matem\'atica,\\
Instituto Superior T\'ecnico, Lisboa, Portugal}\\
{\tt gcardoso@math.ist.utl.pt}\\[1mm]
$^b${\em Institute for Theoretical Physics},
  \\ {\em Utrecht University, Utrecht, The Netherlands}\\
{\tt  B.deWit@uu.nl} \\[1mm]
$^c${\em Nikhef Theory Group, Amsterdam, The Netherlands}\\[1mm]
$^d${\em Physics Department, Utkal University, 
Bhubaneswar 751 004, India}\\
{\tt swapna@iopb.res.in}\\[1mm]

\end{center}
\vskip .2in
\begin{center} {\bf ABSTRACT } \end{center}
\begin{quotation}\noindent 
  The Hesse potential is constructed for a class of four-dimensional
  $N=2$ supersymmetric effective actions with S- and T-duality by
  performing the relevant Legendre transform by iteration. It is a
  function of fields that transform under duality according to an
  arithmetic subgroup of the classical dualities reflecting the
  monodromies of the underlying string compactification. These
  transformations are not subject to corrections, unlike the
  transformations of the fields that appear in the effective action
  which are affected by the presence of higher-derivative couplings.
  The class of actions that are considered includes those of the FHSV
  and the STU model. We also consider heterotic $N=4$ supersymmetric
  compactifications. The Hesse potential, which is equal to the free
  energy function for BPS black holes, is manifestly duality
  invariant. Generically it can be expanded in terms of powers of the
  modulus that represents the inverse topological string coupling
  constant, $g_\mathrm{s}$, and its complex conjugate. The terms
  depending holomorphically on $g_\mathrm{s}$ are expected to
  correspond to the topological string partition function and this
  expectation is explicitly verified in two cases. Terms
  proportional to mixed powers of $g_\mathrm{s}$ and $\bar
  g_\mathrm{s}$ are in principle present.
\end{quotation}

\vfill
\end{titlepage}
\eject
\section{Introduction}
\label{sec:introduction}
\setcounter{equation}{0}
Higher-order curvature corrections to four-dimensional $N=2$
supersymmetric Wilsonian actions are known to affect the duality
transformations of the moduli fields. The full transformation rules
turn out to be much more complicated than their counterpart at the
two-derivative level.  An additional complication arises at the level
of the associated 1PI effective action, where also non-holomorphic
terms need to be incorporated in order to obtain physical results that
reflect the duality invariance. These issues have been discussed in
the context of four-dimensional $N=2$ BPS black holes
\cite{LopesCardoso:2006bg,Cardoso:2008fr}, where it was described how
to incorporate non-holomorphic corrections into the free energy of BPS
black holes in the presence of a Weyl background. This free energy
turns out to be given by the generalized Hesse potential. Unlike the
effective action, the Hesse potential is defined in terms of variables
whose duality transformations are not subject to the deformations
induced by higher-derivative couplings. The relation between the Hesse
potential and the effective action involves a Legendre transform. The
Hesse potential can be regarded as the 'Hamiltonian' version of the
effective Lagrangian, and is invariant under possible duality
transformations. This is comparable to the generic situation for
(abelian) gauge theories and electric/magnetic dualities, where the
Lagrangian is in general not invariant under these dualities, while
the Hamiltonian is invariant. Also the Lagrangian and the Hamiltonian
are related by a Legendre transform and they are expressed in terms of
different dynamical variables.

It is suggestive to assume that the Hesse potential is directly
related to the partition function of the topological string
\cite{Bershadsky:1993cx}. The moduli of the topological string
partition function correspond to moduli of the underlying Calabi-Yau
moduli space and their duality transformations are defined in terms of
monodromy transformations of the Calabi-Yau period vector. These
transformations are thus directly related to the transformations found
at the two-derivative level of the effective action. Therefore, both
the topological string partition function and the Hesse potential are
expressed in terms of variables that transform identically under the
dualities, and moreover, both are duality invariant. In addition, it
has been established that certain string amplitudes are related to the
twisted partition functions of the topological string
\cite{Antoniadis:1993ze,Bershadsky:1993cx}. String amplitudes
correspond to connected field theory graphs, and therefore these
amplitudes must be encoded in both the corresponding effective action
and in the Hesse potential. Consequently the topological string is
contained in the Hesse potential. This leaves the possibility that the
Hesse potential contains more information than just the topological
string in view of the fact that its dependence on the topological
string coupling, $g_\mathrm{s}$, which is inversely proportional to
one of the complex moduli, is in principle not holomorphic.

In general it is not possible to carry out the Legendre transform
explicitly at the level of a full effective action.  This paper is
therefore devoted to carrying out the Legendre transform by iteration
in order to subsequently study the possible relation between the Hesse
potential and the partition function of the topological string. Hence
the BPS black holes will only play an ancillary role in this work. The
Legendre transform leads to new variables for the Hesse potential
which, under duality, transform precisely like the fields used in the
topological string. Unfortunately only in a few cases exact
expressions are known for the topological string, and for the
effective action there is even less data. Consequently we will have to
rely on a restricted number of models with a high degree of
symmetry. However, in principle, our results will also be relevant for
models without duality symmetries.

For a specific class of $N=2$ models, which includes the FHSV
\cite{Ferrara:1995yx} and the STU model
\cite{Sen:1995ff,Gregori:1999ns}, we explicitly compute the Hesse
potential in terms of these new variables, up to second order in the
Weyl background. In the context of the FHSV model, we show that the
Legendre transform reproduces the associated non-holomorphic genus-$2$
partition function of the topological string \cite{Grimm:2007tm},
starting from the expressions found in \cite{Cardoso:2008fr}. In
addition, we consider $N=4$ supersymmetric models in this $N=2$
description, for which we obtain more detailed information on the  
higher-order contributions to the Hesse potential.

Some time ago it has been argued \cite{LopesCardoso:2006bg} that the
exponent of the Hesse potential appears in a fully duality invariant
extension of the OSV integral \cite{Ooguri:2004zv}. A semiclassical
evaluation of this extension reproduces the original OSV integral with
an additional measure factor. At the semiclassical level this modified
integral correctly reproduces all known results for (large) black holes
from both macroscopic and microscopic perspectives.  Beyond the
semiclassical approximation the role of this integral has not been
fully established as yet.

The results of this paper clarify a number of issues in the relation
between the effective action and the Hesse potential. The latter
depends on a modulus that corresponds to the inverse topological
string coupling constant $g_\mathrm{s}$ and on its its complex
conjugate. As it turns out, the topological string partition functions
are recovered when restricting to those terms in the Hesse potential
that depend holomorphically on $g_\mathrm{s}$, at least for genus
$g\leq2$. This sector of the Hesse potential is separately consistent
with respect to duality. In view of the earlier discussion this result
is not unexpected, but the present lack of data on the
higher-derivative terms in the effective actions forms an obstacle for
uncovering the more conceptual aspects of the relation between the
Hesse potential and the topological string partition function. In
principle, the Hesse potential will also contain terms proportional to
mixed powers of $g_\mathrm{s}$ and $\bar g_\mathrm{s}$. It is
important to realize that these are not primarily induced by the
non-holomorphic corrections associated with non-local terms in the
effective action, but they are present as a result of the Legendre
transform. We observe that, at genus 2, these mixed terms can be
absorbed by suitable contributions from the effective action. Without
detailed knowledge of the latter, it is not clear how to establish
this in some generality, especially because the consequences of
(partially) absorbing these terms can only be seen at higher
genus. Perhaps the analysis can be strengthened eventually by taking
into account information on the asymptotic behaviour of the functions
involved. Admittedly, the situation remains rather complex, but it is
clear that considerable progress can be made on the basis of the case
studies considered in this paper.

This paper is organized as follows.  In section
\ref{sec:Hesse-potential}, after reviewing the construction of the
Hesse potential in the presence of non-holomorphic terms, we introduce
new variables ${\tilde Y}^I$ for the Hesse potential that transform
under duality according to the classical transformation rules.
Section \ref{sec:S-T-dualities} contains a brief review of the
consequences of S- and T-duality invariance for a class of $N=2$
models that contain the FHSV and the STU model. In section
\ref{sec:perf-legendre-transf} we construct the variables $\tilde Y^I$
for this class of $N=2$ models.  We derive a set of equations that
these new variables have to satisfy, and we solve them iteratively in
the Weyl background, up to second order. We verify that these new
parameters satisfy the required duality properties. Section
\ref{sec:heterotic-example} deals with models corresponding to
heterotic $N=4$ compactifications, for which we derive all-order
results. In section \ref{sec:hesse-potential-2-order} we compute the
Hesse potential for the more generic case, expressed in the new
variables, to second order. Section \ref{sec:hesse-specific}
summarizes the situation for specific models and compares the results
to the twisted partition functions of the topological string. 
In two cases we demonstrate that these 
partition functions can be reproduced by the
corresponding terms in the Hesse potential 
that depend holomorphically on $g_\mathrm{s}$. 

\section{The Hesse potential}
\label{sec:Hesse-potential}
\setcounter{equation}{0}
For $N=2$ supergravity the part of the Lagrangian pertaining to the
vector supermultiplets is encoded in a holomorphic function $F$ of the
complex scalar fields $X^I$ belonging to these multiplets, which is
homogeneous of second degree. Here the index $I=0,1,\ldots,n$ labels
the various vector multiplets. The vector multiplets have an optional
coupling to the square of the Weyl tensor, which can be encoded in the
function $F$ by introducing a dependence on another complex scalar
field equal to the square of the anti-selfdual antisymmetric auxiliary
field that constitutes the lowest-weight field of the so-called
Weyl supermultiplet.  All these scalars are defined projectively, but
in the context of BPS black holes suitably normalized fields have been
introduced denoted by $Y^I$ and $\Upsilon$
\cite{LopesCardoso:1998wt}. In terms of these fields the attractor
equations for BPS black holes take the form,
\begin{equation}
  \label{eq:BPS-attractor}
  Y^I-\bar Y^I = \mathrm{i} p^I\,,\qquad F_I-\bar F_I= \mathrm{i}
  q_I\,,\qquad \Upsilon= -64\,,
\end{equation}
where $F_I$ denotes the derivative of the function $F$ with respect to
$Y^I$ and $\bar F_I$ is its complex conjugate.\footnote{
  Hence $\bar F_I$ equals the derivative of $\bar F$ with respect to
  $\bar Y^I$. We refrain from distinguishing holomorphic and
  anti-holomorphic derivatives, $\partial/\partial Y^I$ and
  $\partial/\partial\bar Y^I$, by the use of different types of
  indices. } 
The first two atractor equations can be obtained from extremizing the
BPS free energy. This observation will be relevant in the sequel. 

The above does not yet account for the presence of non-holomorphic
modifications. These modifications signal departures from the
Wilsonian action that originate from integrating out the massless
modes in order to obtain the full effective action. This integration
gives rise to non-local terms in the corresponding supergravity
action.  Unfortunately not much is known about these non-localities,
except that they are often required to preserve physical symmetries
that cannot be fully realized at the level of the Wilsonian action.
An early example of this phenomenon can be found in
\cite{Dixon:1990pc}, where it was demonstrated that the gauge coupling
constants in heterotic string compactifications are moduli dependent
with non-holomorphic corrections.  Also in the context of BPS black
holes the need for non-holomorphic modifications has been demonstrated
to ensure that the `period vector' $(Y^I,F_I)$ transforms consistently
under S-duality \cite{LopesCardoso:1999ur}.  When these modifications
are taken into account an S-duality invariant entropy is obtained. The
results of this analysis are in accord with the results for the
non-holomorphic terms found in the corresponding effective action
\cite{Harvey:1996ir}. More recently, it has been shown
\cite{Cardoso:2004xf,Jatkar:2005bh} how the same results emerge from a
semiclassical approximation of the microscopic degeneracy formula for
$N=4$ dyons \cite{Dijkgraaf:1996it,Shih:2005uc,Jatkar:2005bh,
  David:2006yn,Banerjee:2008pv, Banerjee:2008pu,Dabholkar:2008zy}.

In order to ensure that the attractor equations will still follow from
a variational principle in the presence of non-holomorphic
corrections, it turns out that these corrections must be encoded in a
real and homogeneous function of second degree denoted by
$\Omega(Y,\bar Y,\Upsilon,\bar\Upsilon)$, which is incorporated into
the function $F$ in the following way
\cite{Cardoso:2004xf,Cardoso:2008fr},
\begin{equation}
  \label{eq:F-decomposition}
  F = F^{(0)}(Y,\Upsilon) + 2\mathrm{i}\,\Omega(Y,\bar
  Y,\Upsilon,\bar\Upsilon) \,.
\end{equation}
The attractor equations \eqref{eq:BPS-attractor} retain the same form,
irrespective of the presence of these non-holomorphic terms. Although
the explicit couplings in the Lagrangian corresponding to this
modification are unknown, it turned out that important progress can be
made without first constructing the full effective action.  When the
function $\Omega$ is harmonic, i.e., when it can be written as the sum
of a holomorphic and an anti-holomorphic function, then one may simply
absorb the holomorphic part into the first term. The anti-holomorphic
part will then not contribute as it will vanish under the holomorphic
derivatives which enter the attractor equations as well as the black
hole entropy. Consequently we can incorporate the $\Upsilon$-dependent
terms in $F^{(0)}$ into $\Omega$. In that case $F^{(0)}(Y)$ will no
longer depend on $\Upsilon$, and will refer to the {\it
  classical} contribution that pertains to the part of the Lagrangian
quadratic in space-time derivatives.

In this context there exists the notion of a BPS free energy, which is
defined as follows \cite{LopesCardoso:2006bg},
\begin{equation}
  \label{eq:free-energy-phase}
  \mathcal{F}(Y,\bar Y,\Upsilon,\bar\Upsilon)= - \mathrm{i} \left(
  {\bar Y}^I F_I - Y^I {\bar F}_I
  \right) - 2\mathrm{i} \left( \Upsilon F_\Upsilon - \bar \Upsilon
  \bar F_{\bar\Upsilon}\right)\,,
\end{equation}
where $F_\Upsilon= \partial F/\partial\Upsilon$. We recall that each
of the two terms in (\ref{eq:free-energy-phase}) transform as a
function under electric/magnetic duality. This free energy, whose
existence seems desirable based on semiclassical arguments, enters the
BPS entropy function $\Sigma$, defined by
\begin{equation}
  \label{eq:Sigma-simple}
  \Sigma(Y,\bar Y,p,q) =  \mathcal{F}(Y,\bar Y)
  - q_I   (Y^I+\bar Y^I)   + p^I (F_I+\bar F_I)  \;, 
\end{equation}
where the black hole charges $q_I$ and $p^I$ couple to the
corresponding electro- and magnetostatic potentials at the horizon,
which are equal to \cite{LopesCardoso:2000qm},
\begin{equation}
  \label{eq:real-special-vars}
  \phi^I = Y^I + \bar Y^I\,,\qquad \chi_I = F_I + \bar F_{I}\,. 
\end{equation}
In \cite{Cardoso:2008fr} we have indicated how these expressions are
consistent with electric/magnetic duality. Requiring stationarity of
the entropy function $\Sigma$ with respect to $Y^I$ leads directly to
the attractor equations \eqref{eq:BPS-attractor} while the value of
$\Sigma$ at the attractor point defines the macroscopic
(field-theoretic) entropy divided by $\pi$. In the absence of
non-holomorphic modifications, this entropy has been shown
\cite{LopesCardoso:1998wt} to coincide with Wald's entropy based on a
Noether charge \cite{Wald:1993nt,Jacobson:1993vj,Iyer:1994ys}. Under
electric/magnetic duality the charges $(p^I,q_I)$ and the `period
vector' $(Y^I, F_I)$ transform under symplectic (real) rotations, and
so do the potentials $(\phi^I,\chi_I)$.

In the presence of higher-order derivative actions in the effective
action, the original complex fields $Y^I$ transform in a complicated
way under electric/magnetic duality. Therefore it is advantageous to
consider a variable change to the real coordinates $\phi^I$ and
$\chi_I$, which transform linearly under dualities. This conversion is
well defined whenever $\det[F_{IJ} - \bar F_{IJ}]\not=0$, where
$F_{IJ}$ denotes the second derivative of $F$ with respect to $Y^I$
and $Y^J$. As we shall demonstrate shortly, the so-called Hesse
potential, defined as the Legendre transform of a linear combination of
the imaginary part of $F$ and $\Omega$ with respect to the imaginary
part of the $Y^I$, is a function of $\phi^I$ and $\chi_I$. It is a
generalization of the Hesse potential defined in the context of
special geometry \cite{Freed:1997dp,Alekseevsky}. To perform the
conversion to real variables $\phi^I$ and $\chi_I$, we first decompose
$Y^I$ and $F_I$ into their real and imaginary parts,
\begin{equation}
  \label{eq:u-y}
  Y^I = \tfrac12(\phi^I + \mathrm{i} u^I) \;,\qquad F_I =
  \tfrac12(\chi_I + \mathrm{i} v_I)  \;. 
\end{equation}
The real parametrization is obtained by taking $(\phi^I,\chi_I,
\Upsilon, {\bar \Upsilon})$ instead of $(Y^I, {\bar Y}^I, \Upsilon,
{\bar \Upsilon})$ as the independent variables. Although $\Upsilon$ is
a spectator, note that the inversion of $\chi_I =
\chi_I(\phi,u,\Upsilon,{\bar \Upsilon})$ gives ${\rm Im}\; Y^I =
u^I(\phi,\chi,\Upsilon, {\bar \Upsilon})$. To compare partial
derivatives in the two parametrizations, we need,
\begin{eqnarray}
  \label{eq:derivative-u,y}
  \frac{\partial}{\partial\phi^I}\Big\vert_u &=& \frac{\partial}{\partial
  \phi^I}\Big\vert_\chi + \frac{\partial
  \chi_J(\phi,u,\Upsilon,\bar\Upsilon)}{\partial\phi^I} \, 
  \frac{\partial}{\partial \chi_J}\Big\vert_\phi\;, \nonumber\\
  \frac{\partial}{\partial u^I}\Big\vert_\phi &=& \frac{\partial
  \chi_J(\phi,u,\Upsilon,\bar\Upsilon)}{\partial u^I} \,
  \frac{\partial}{\partial\chi_J}\Big\vert_\phi\;, \nonumber\\
  \frac{\partial}{\partial \Upsilon}\Big\vert_{\phi,u} &=&
  \frac{\partial}{\partial \Upsilon}\Big\vert_{\phi,\chi} + \frac{\partial
  \chi_I(\phi,u,\Upsilon,\bar\Upsilon)}{\partial \Upsilon} \,
  \frac{\partial}{\partial\chi_I}\Big\vert_\phi\;.
\end{eqnarray}
The homogeneity is preserved under the reparametrization because
$\chi(\phi,u,\Upsilon,\bar\Upsilon)$ is a homogeneous function of
first degree. This results in the equality,
\begin{eqnarray}
  \label{eq:homogeneity-u,y}
  &&
  \phi^I\,\frac{\partial}{\partial \phi^I}\Big\vert_u+
  u^I\,\frac{\partial}{\partial u^I}\Big\vert_\phi +
  2\, \Upsilon\,\frac{\partial}{\partial \Upsilon}\Big\vert_{\phi,u} +
  2\, \bar\Upsilon\,\frac{\partial}{\partial
  \bar\Upsilon}\Big\vert_{\phi,u} \nonumber\\
&&{}=
  \phi^I\,\frac{\partial}{\partial \phi^I}\Big\vert_\chi+
  \chi_I\,\frac{\partial}{\partial \chi_I}\Big\vert_\phi +
  2\, \Upsilon\,\frac{\partial}{\partial \Upsilon}\Big\vert_{\phi,\chi} +
  2\, \bar\Upsilon\,\frac{\partial}{\partial
  \bar\Upsilon}\Big\vert_{\phi,\chi}\;.
\end{eqnarray}

The Hesse potential is defined as the Legendre transform of
$4(\mathrm{Im} \, F- \Omega)$ with respect to
$u^I=2\,\mbox{Im}\,Y^I$,\footnote{
  See \cite{LopesCardoso:2006bg}; note that the conventions of this
  paper are not the same.} 
\begin{equation}
  \label{eq:GenHesseP}
  \mathcal{H}(\phi,\chi,\Upsilon, {\bar \Upsilon}) = 4 \; {\rm
  Im}\,F(Y,\bar Y,\Upsilon,\bar\Upsilon) -4\,\Omega(Y,\bar
  Y,\Upsilon,\bar\Upsilon)  -   \chi_I \,u^I \;, 
\end{equation}
which is a homogeneous function of second degree. Note that
$\delta\mathcal{H} = v_I\,\delta\phi^I - u^I\,\delta\chi_I$, which
shows that the attractor equations \eqref{eq:BPS-attractor} take the
form,
\begin{equation}
  \label{eq:ExtremEqs}
  \frac{\partial \mathcal{H}}{\partial \phi^I} = q_I \;,\qquad
  \frac{\partial \mathcal{H}}{\partial \chi_I} = - p^I  \;.
\end{equation}
These equations follow from requiring that the entropy function 
\begin{equation}
  \label{eq:RealSigma}
  \Sigma(\phi,\chi,p,q) =\mathcal{H}(\phi,\chi,\Upsilon,\bar\Upsilon) -
  q_I \,\phi^I + p^I\, \chi_I \;,
\end{equation}
is stationary. Comparing this result to the entropy function
\eqref{eq:Sigma-simple} indicates that the Hesse potential is just the
BPS free energy \eqref{eq:free-energy-phase}. Indeed, using the
homogeneity properties of $F$ and $\Omega$, we establish the relation,
\begin{equation}
  \label{eq:GenHesseP2}
  \mathcal{H}(\phi,\chi,\Upsilon, {\bar \Upsilon}) = - \mathrm{i}( {\bar 
  Y}^I F_I - Y^I {\bar F}_I  )
  - 2\,\mathrm{i} (\Upsilon F_\Upsilon - {\bar \Upsilon} {\bar F}_{{\bar
    \Upsilon}}) = \mathcal{F}(Y,\bar Y,\Upsilon,\bar \Upsilon)  \;.
\end{equation}
Substituting the result of the attractor equations into the entropy
function thus yields the macroscopic BPS entropy, just as before
(irrespective of the non-holomorphic modification).  In the spirit of
\cite{Ooguri:2004zv} it has been proposed that the integral over
$\exp[\Sigma(\phi,\chi,p,q)]$ yields the entropy for BPS black holes,
and this proposal has been verified in a variety of cases at the
semiclassical level \cite{LopesCardoso:2006bg}. Some of its
implications have also successfully been confronted with microscopic
counting data, mainly from heterotic $N=4$ supersymmetric models
\cite{Dijkgraaf:1996it,Cardoso:2004xf,Shih:2005uc,Shih:2005he,Jatkar:2005bh}.\footnote{
  For an $N=2$ application, see \cite{Cardoso:2008ej}.
} 
However, we should note that no exact results are available as yet. 

Under duality invariances the complex variables $Y^I$ transform in a
complicated way, which can be studied order-by-order in $\Upsilon$
\cite{Cardoso:2008fr}. To explicitly establish the invariance of the
BPS free energy is thus cumbersome, as both the transformation rules
and the expression for the free energy take the form of a power series
in $\Upsilon$. This was analyzed extensively in \cite{Cardoso:2008fr},
where arguments were put forward that show that the duality invariance
persists in the presence of the non-holomorphic modifications. The
Hesse potential depends on fields $(\phi^I,\chi_I)$ that transform
under the dualities with a real symplectic rotation, just as the
charges $(p^I,q_I)$.  These rotations, referred to as monodromies, are
fixed from the start and cannot be subject to any iterative procedure
(because of the integer-valued charge lattice).  Therefore, it is in
principle easier to consider the duality invariance of the Hesse
potential, but this quantity has to be evaluated by a Legendre
transform which cannot be explicitly performed and requires an
iterative procedure.

In this paper we will evaluate the first few terms of the expansion of
the Hesse potential in terms of $\Upsilon$ for a class of $N=2$
effective actions with S- and T-duality.  Subsequently, we will verify
the duality invariance of the terms in the expansion and compare them
to results known for the topological string partition function. It is
rather convenient to do this in terms of different variables than the
real potentials $(\phi^I,\chi_I)$.  Namely, we will re-express the
$(\phi^I,\chi_I)$, which incorporate all the terms of the action, in
terms of new complex fields denoted by $\tilde Y^I$, which will
coincide precisely with the fields $Y^I$ that one would obtain from
$(\phi^I,\chi_I)$ by using only the lowest-order holomorphic function
$F^{(0)}$.  Hence the identification proceeds as follows,
\begin{eqnarray}
  \label{eq:tilde-fields-Y}
  2\,\mathrm{Re}\, Y^I &=& \phi^I ~=~ 2\,\mathrm{Re}\,\tilde Y^I
  \,,\nonumber\\ 
  2\,\mathrm{Re}\,F_I(Y,\bar Y,\Upsilon,\bar\Upsilon) &=& \chi_I ~=~
  2\,\mathrm{Re}\,F_I^{(0)}(\tilde Y) \,. 
\end{eqnarray}
At the classical level $\tilde Y^I=Y^I$, but in higher orders the
relation between these moduli is complicated and will depend on
$\Upsilon$. The crucial point is that the duality transformations for
the fields $\tilde Y^I$ will be independent of $\Upsilon$ and its
complex conjugate, unlike the transformations of the fields $Y^I$,
which depend non-trivially on $\Upsilon,\bar\Upsilon$. Therefore, the
moduli $\tilde Y^I$ are expected to be the appropriate variables for
the topological string. We note that the passage from the supergravity
(or effective action) variables $Y^I$ to the topological string
variables ${\tilde Y}^I$ induces a change of complex structure (which
is thus not primarily induced by the non-holomorphic terms contained
in the function $\Omega$). This will become evident in section
\ref{sec:perf-legendre-transf}, where we compute the change for a
class of models with a high degree of symmetry. Observe that the
left-hand side of the second equation \eqref{eq:tilde-fields-Y}
depends explicitly on the function $\Omega$, whereas the relation
between $(\phi,\chi)$ and the $\tilde Y^I$ represents a simple change
of variables. As it turns out, this change of variables facilitates
the calculations that we will perform in later sections.

It is easy to verify that the covariant moduli proposed for the STU
model in \cite{Banerjee:2009am} do not fall in the same class as the
moduli $\tilde Y^I$, simply because they do not satisfy
\eqref{eq:tilde-fields-Y}. It remains to be seen what the relation
between the two sets of covariant variables implies. At any rate, the
variables used in this paper exist generally, outside the context of a
specific model.

\section{S- and T-dualities}
\label{sec:S-T-dualities}
\setcounter{equation}{0}
Following \cite{Cardoso:2008fr} we consider a class of models for
which the lowest-order contribution of the action is encoded in the
holomorphic, homogeneous function,
\begin{equation}
  \label{eq:F-0}
  F^{(0)}(Y) = - \frac{Y^1 Y^a\eta_{ab}Y^b}{Y^0}  \,, 
\end{equation}
where $a,b=2,\ldots,n$, and the symmetric matrix $\eta_{ab}$ is an
$\mathrm{SO}(n-2,1)$ invariant metric of indefinite signature. To this
expression we will add the homogeneous real function $\Omega$ as
specified in \eqref{eq:F-decomposition}, which, as explained in the
previous section, encodes certain higher-order derivative as well as
non-local interactions. Models of this type arise in type-II
compactifications on Calabi-Yau three-folds that are K3 fibrations.
The number $n$ depends on the particular model that one is
considering. Both the FHSV \cite{Ferrara:1995yx} and the STU model
\cite{Sen:1995ff,Gregori:1999ns} belong to this class and have $n=11$
and $n=3$, respectively. For these models explicit information is
available for the terms of higher-order in $\Upsilon$
\cite{Grimm:2007tm,Cardoso:2008fr}. Furthermore we use the $N=2$
supergravity description to also consider a number of heterotic string
compactifications with $N=4$ supersymmetry.

In the absence of non-holomorphic corrections the function
$F(Y,\Upsilon)$ takes the form of a loop expansion with $Y^0$ as a
loop-counting parameter,
\begin{equation}
  \label{eq:top-string}
  F(Y,\Upsilon) =  \mathrm{i} (Y^0)^2 \, S\,T^a\eta_{ab}T^b +
  \Upsilon\, F^{(1)}(S,T) + \sum_{g=2}^\infty\,
  \frac{\Upsilon^g}{(Y^0)^{2g-2}} \, F^{(g)}(S,T)  \,,
\end{equation}
with `special coordinates' defined in the usual fashion,
\begin{equation}
  \label{eq:special-coordinates}
  S=-\mathrm{i}\,Y^1/Y^0\,,\qquad T^a=-\mathrm{i}\,Y^a/Y^0\,.
\end{equation}
In the context of type-II models based on K3-fibered Calabi-Yau
three-folds, these special coordinates can be used to parametrize
(half of) the moduli space of the associated string compactification.
An expansion such as \eqref{eq:top-string} is also relevant for the
topological string on the same Calabi-Yau three-fold, where $Y^0$ is
regarded as the inverse topological string coupling constant and the
functions $F^{(g)}(S,T)$ are the genus-$g$ twisted partition
functions.\footnote{
  Hence $ F^{(g)}(Y) = (Y^0)^{2-2g} \, F^{(g)}(S,T)$; when referring
  to the genus-$g$ partition functions in the text, we usually do not
  make a distinction between $F^{(g)}(Y)$ and $F^{(g)}(S,T)$. 
  \label{eq:FgY-Fg}
} 
The latter acquire non-holomorphic corrections encoded in the
holomorphic anomaly equation, whose structure is such that the
holomorphic dependence on the topological string coupling constant is
preserved \cite{Bershadsky:1993cx}. As we already mentioned,
non-holomorphic corrections are also required to realize the relevant
symmetries of the effective action \cite{Dixon:1990pc}, and it seems
likely that these two phenomena are in fact related. To clarify this
in some detail was in fact one of the motivations for the work
described here. However, we should stress that in spite of the fact
that the same expansion \eqref{eq:top-string} is relevant for both the
effective action and for the topological string, the two expansions
should not be identified, as was demonstrated in
\cite{Cardoso:2008fr}. In the next section we will work out the
precise correspondence in more detail by explicitly performing the
Legendre transform for the Hesse potential, up to $g=2$.

In this section we will be reviewing the S- and T-duality
transformations based on \cite{Cardoso:2008fr}, where the effect of the
dualities was determined for the partial derivatives of $\Omega$. For
the models based on \eqref{eq:F-0} the duality group is given by
$\mathrm{SL}(2;\mathbb{Z})\times\mathrm{O}(n-1,2;\mathbb{Z})$, where
the first factor refers to the S-duality group and the second one to
the T-duality group. When including the $\Upsilon$-dependent terms
according to \eqref{eq:F-decomposition}, only a subgroup may be
realized. In \cite{Cardoso:2008fr} the requirements for the function
$\Omega$ were derived based on the assumption that the invariance was
realized for a suitable arithmetic subgroup. Under this group the
$(Y^I,F_I)$ transform as follows under S-duality,
\begin{equation}
    \label{eq:S-duality}
    \begin{array}{rcl}
      Y^0 &\to& d \, Y^0 + c\, Y^1 \;,\\
      Y^1 &\to& a \, Y^1 + b \, Y^0 \;,\\
      Y^a &\to& d\, Y^a - \ft12 c \,\eta^{ab}\,F_b  \;,
    \end{array}
    \quad
    \begin{array}{rcl}
      F_0 &\to& a\,  F_0 -b\,F_1 \;, \\
      F_1 &\to& d \,F_1 -c\, F_0 \;,\\
      F_a &\to& a \, F_a -2b\, \eta_{ab}\, Y^b \;,
    \end{array}
\end{equation}
where $a,b,c,d$ are integer-valued parameters that satisfy $ad-bc=1$
which parametrize (a subgroup of) $\mathrm{SL}(2;\mathbb{Z})$.

For the T-duality group, general transformations are conveniently
generated by products of a number of specific finite transformations.
Those belonging to the $\mathrm{O}(n-2,1;\mathbb{Z})$ subgroup are
manifest in the above description and do not need to be considered.
Then there are $n-1$ abelian transformations generated by
\begin{equation}
  \label{eq:shifs}
    \begin{array}{rcl}
      Y^0 &\to& Y^0 \,,\\
      Y^1 &\to& Y^1 \,,\\
      Y^a &\to& Y^a -\lambda^a \,Y^0\,,
    \end{array} 
    \quad
    \begin{array}{rcl}
      F_0 &\to& F_0 +\lambda^a F_a + \lambda^a\eta_{ab}\lambda^b \,Y^1
      \,,\\ 
      F_1 &\to& F_1 +2\,\lambda^a \eta_{ab} Y^b -
      \lambda^a\eta_{ab}\lambda^b \,Y^0 \,,\\
      F_a &\to& F_a +2\,\eta_{ab}\lambda^b \,Y^1 \,,
  \end{array}
\end{equation}
where the $\lambda^a$ are integers. The full
$\mathrm{O}(n-1,2;\mathbb{Z})$ group is generated upon including also
the following transformation,
\begin{equation}
    \label{eq:T-inversion}
    \begin{array}{rcl}
      Y^0 &\to& F_1 \;,\\
      Y^1 &\to& - F_0  \;,\\
      Y^a &\to&  Y^a  \;,
    \end{array}
    \quad
    \begin{array}{rcl}
      F_0 &\to& -Y^1 \;, \\
      F_1 &\to& Y^0 \;, \\
      F_a &\to& F_a \;,
    \end{array}
\end{equation}
which squares to the identity.

In the case that $\Omega$ is suppressed in \eqref{eq:F-decomposition},
it is straightforward to evaluate the behaviour of these
transformations on the special coordinates $S$ and $T^a$, and on the
remaining field $Y^0$. Under S-duality we find,
\begin{equation}
  \label{eq:ST-Sdual}
  S \rightarrow
  \frac{a\,S-\mathrm{i}b}{d+ \mathrm{i}c\,S} \;, 
  \qquad T^a\to T^a\;\qquad Y^0\to (d+ \mathrm{i}c\,S)\, Y^0 \; .
\end{equation}
The T-duality transformations \eqref{eq:shifs} and
\eqref{eq:T-inversion} lead to, respectively,
\begin{equation}
  \label{eq:T-S-Sdual}
  S\to S\;, \qquad T^a\to T^a+\mathrm{i}
  \,\lambda^a \;, \qquad T^a\to 
  \frac{T^a}{T^b\eta_{bc}T^c}\;\qquad Y^0\to
  T^b\eta_{bc}T^c \,Y^0 \; . 
\end{equation}
These S- and T-duality transformations become much more complicated
when $\Omega$ is taken into account in \eqref{eq:F-decomposition}.
Insisting on the same symmetry (i.e., characterized by the same
transformations acting on $(Y^I,F_I)$), or a subgroup thereof,
severely restricts the $\Upsilon$-dependent contributions contained
in $\Omega$. These restrictions take the form of prescribed
transformation rules for the first-order derivatives $\Omega$ with
respect to the fields. The crucial observation is, however, that the
fields $\tilde S$, $\tilde T^a$ and $\tilde Y^0$, based on
\eqref{eq:tilde-fields-Y}, with
\begin{equation}
  \label{eq:special-coordinates-tilde}
  \tilde S=-\mathrm{i}\,\tilde Y^1/\tilde Y^0\,,\qquad
  \tilde T^a=-\mathrm{i}\,\tilde Y^a/\tilde Y^0\,, 
\end{equation}
will still transform exactly as in \eqref{eq:ST-Sdual} and
\eqref{eq:T-S-Sdual}. Of course, this is true provided the
$\Upsilon$-dependent terms satisfy the correct symmetry properties.
These will be summarized below.

Since the S- and T-duality transformations involve the derivatives
$F_I$, we note the expressions, 
\begin{eqnarray}
  \label{eq:F_I}
  F_0&=& \frac{Y^1}{(Y^0)^2}\,Y^a\eta_{ab}Y^b -\frac{2\mathrm{i}}{Y^0}
  \left[-Y^0 \frac{\partial}{\partial Y^0}+
  S\frac{\partial}{\partial S}+ T^a\frac{\partial}{\partial T^a}
  \right]\Omega \;,  \nonumber\\ 
  F_1&=&{}- \frac{1}{Y^0}\,Y^a\eta_{ab}Y^b +\frac{2}{Y^0} \,
  \frac{\partial\Omega}{\partial S} \;,\nonumber\\
  F_a&=& {}- 2 \frac{Y^1}{Y^0}\,\eta_{ab}Y^b +\frac{2}{Y^0} \,
  \frac{\partial\Omega}{\partial T^a} \;,
\end{eqnarray}
where we regard $\Omega$ as a function of $Y^0$, $S$ and $T^a$ (and
their complex conjugates). 

With these results the S-duality transformations \eqref{eq:S-duality}
take the form,
\begin{eqnarray}
  \label{eq:full-S}
  Y^0&\to& \Delta_{\mathrm{S}}\, Y^0\;, \nonumber\\
  Y^1&\to& a\,Y^1+ b\,Y^0\;, \nonumber\\
  Y^a&\to& \Delta_{\mathrm{S}}\, Y^a - \frac{c}{Y^0} \,\eta^{ab}\,
  \frac{\partial\Omega}{\partial T^b}  \;,  
\end{eqnarray}
with 
\begin{equation}
  \label{eq:Delta-S}
  \Delta_{\mathrm{S}} = d+\mathrm{i}c\,S\,.
\end{equation}
On the special coordinates $S$ and $T^a$ these transformations extend
the previous result \eqref{eq:ST-Sdual},
\begin{equation}
  \label{eq:ST-S-full}
  S \rightarrow
  \frac{a\,S-\mathrm{i}b}{\mathrm{i}c\,S+d} \;, 
  \qquad T^a\to T^a +\frac{\mathrm{i} c}{\Delta_{\mathrm{S}}\,(Y^0)^2}
  \,\eta^{ab} 
  \,\frac{\partial\Omega}{\partial T^b}  \;,
\end{equation}
and we note the useful relations
\begin{equation}
  \label{eq:dS-dS}
  \frac{\partial{S^\prime}}{\partial{S}}
  =\Delta_{\mathrm{S}}{}^{-2}\,,\qquad \frac1{S+\bar S}\to
  \frac{\vert\Delta_{\mathrm{S}}\vert^2}{S+\bar S} = 
  \frac{\Delta_{\mathrm{S}}{}^2}{S+\bar S}
  -\mathrm{i} c\,\Delta_{\mathrm{S}}\,.  
\end{equation}
Assuming that the above transformations constitute an invariance of
the model, we require that the S-duality transformations of the $Y^I$
induce the expected transformations of the $F_I$ upon substitution.
This leads to the following result,\footnote{
$(O)^\prime_{\mathrm{S,T}}$
denotes the change of $O$ under S- or T-duality induced by the 
transformation of all the arguments on which $O$ depends.}
\begin{eqnarray}
  \label{eq:S-invariance}
  \left(\frac{\partial\Omega}{\partial T^a}\right)^\prime_\mathrm{S} &=&
  \frac{\partial\Omega}{\partial T^a} \;, \nonumber\\
  \left(\frac{\partial\Omega}{\partial S}\right)^\prime_\mathrm{S} -
  \Delta_{\mathrm{S}}{}^2\,\frac{\partial\Omega}{\partial S} &=&
  \frac{\partial(\Delta_{\mathrm{S}}{}^2)}{\partial{S}} 
  \left[-\tfrac12 Y^0 \frac{\partial\Omega}{\partial Y^0}
  -\frac{\mathrm{i}c}{4\,\Delta_{\mathrm{S}}\,(Y^0)^2}
  \,\frac{\partial\Omega}{\partial T^a}\eta^{ab}
  \frac{\partial\Omega}{\partial T^b} \right] \;,
  \nonumber\\ 
  \left(Y^0\frac{\partial\Omega}{\partial Y^0}\right)^\prime_\mathrm{S} &=&
  Y^0 \frac{\partial\Omega}{\partial Y^0} 
  +\frac{\mathrm{i}c}{\Delta_{\mathrm{S}}\,(Y^0)^2}
  \,\frac{\partial\Omega}{\partial T^a}\eta^{ab}
  \frac{\partial\Omega}{\partial T^b}\;. 
\end{eqnarray}

The same reasoning applies to T-duality. Under the transformation
\eqref{eq:shifs} it follows from \eqref{eq:F_I} that all the
derivatives $\partial\Omega/\partial Y^0$, $\partial\Omega/\partial S$
and $\partial\Omega/\partial T^a$ must be invariant under integer
shifts $T^a\to T^a + \mathrm{i}\,\lambda^a$. 
For the T-duality transformation \eqref{eq:T-inversion} the analysis
is more subtle. Using \eqref{eq:F_I} we derive,
\begin{eqnarray}
  \label{eq:full-T}
  Y^0&\to& \Delta_{\mathrm{T}}\, Y^0\;, \nonumber\\
  Y^1&\to& \Delta_{\mathrm{T}}\, Y^1 + \frac{2\mathrm{i}}{Y^0}
  \left[-Y^0\frac{\partial\Omega}{\partial Y^0} +
  T^a\frac{\partial\Omega}{\partial T^a}\right]\;, \nonumber\\
  Y^a&\to& Y^a \;,
\end{eqnarray}
with 
\begin{equation}
  \label{eq:Delta-T}
  \Delta_{\mathrm{T}} = T^a\eta_{ab}T^b +\frac{2}{(Y^0)^2}
  \frac{\partial\Omega}{\partial S}\;. 
\end{equation}
On the special coordinates the transformation \eqref{eq:full-T}
extends the previous result \eqref{eq:T-S-Sdual},
\begin{eqnarray}
  \label{eq:ST-T-full}
  S&\rightarrow&S+ \frac2{\Delta_{\mathrm{T}}(Y^0)^2}
  \,\left[- Y^0 \frac{\partial\Omega}{\partial Y^0} + 
  T^a\frac{\partial\Omega}{\partial T^a} \right]  \;,
  \nonumber\\ 
  \qquad T^a&\to& \frac{T^a}{\Delta_{\mathrm{T}}}  \;.
\end{eqnarray} 
When the $\partial\Omega/\partial S$ term is suppressed in
\eqref{eq:Delta-T}, one obtains the result,
\begin{equation}
  \label{eq:re-T-square}
  (T+\bar T)^a\eta_{ab}(T+\bar T)^b \to
  \frac{1}{\vert\Delta_\mathrm{T}\vert^{2}} \, 
  (T+\bar T)^a\eta_{ab}(T+\bar T)^b\,. 
\end{equation}
Assuming again that the above transformations constitute an invariance
of the model, so that the T-duality transformation \eqref{eq:full-T}
of the $Y^I$ induces the expected transformations of the $F_I$ upon
substitution, leads to
\begin{eqnarray}
  \label{eq:T-invariance}
  \left(\frac{\partial\Omega}{\partial S}\right)^\prime_\mathrm{T} &=& 
  \frac{\partial\Omega}{\partial S} \;, \nonumber\\
  \left(\frac{\partial\Omega}{\partial T^a}\right)^\prime_\mathrm{T} &=&
  \left(\Delta_{\mathrm{T}} \,\delta_a{}^b - 2\,\eta_{ac}T^c
  T^b\right) \,\frac{\partial\Omega}{\partial T^b}
  +2\,\eta_{ab}T^b\;Y^0 \frac{\partial\Omega}{\partial Y^0} \;,
  \nonumber\\ 
  \left(Y^0 \frac{\partial\Omega}{\partial Y^0}\right)^\prime_\mathrm{T} &=&
  Y^0 \frac{\partial\Omega}{\partial Y^0} + 
  \frac{4}{\Delta_{\mathrm{T}}\,(Y^0)^2}
  \,\frac{\partial\Omega}{\partial S} 
  \left[- Y^0 \frac{\partial\Omega}{\partial Y^0} +
  T^a\frac{\partial\Omega}{\partial T^a}\right]\;. 
\end{eqnarray}

This completes the review of the requirements for the function
$\Omega$ derived in \cite{Cardoso:2008fr}.
We stress
once more that the central results, \eqref{eq:S-invariance} and
\eqref{eq:T-invariance}, hold in the presence of non-holomorphic
modifications. Furthermore, it should be clear that $\Omega$ is not an
invariant function. While the fields $\Upsilon$ and $\bar\Upsilon$ do
not enter explicitly into the monodromies \eqref{eq:S-duality},
\eqref{eq:shifs} and \eqref{eq:T-inversion}, the corresponding
transformations induced on $Y^0$, $S$, and $T^a$ depend in a
complicated way on $\Upsilon$ and $\bar\Upsilon$. 

\section{Performing the Legendre transform}
\label{sec:perf-legendre-transf}
\setcounter{equation}{0}
In this section we will now consider the new variables $\tilde Y^I$,
and the corresponding set of variables consisting of $\tilde Y^0,
\tilde S$ and $\tilde T^a$. To explicitly evaluate the equations that
determine the new variables is a rather laborious task. They involve
polynomials of fourth degree in the various fields which we shall
subsequently solve by iteration. This iteration leads to infinite
expansions, which in most cases we truncate at some order. The reader
who is not primarily interested in these manipulations, may skip this
section upon first reading. At the end of the section we also
reconsider the S- and T-duality transformations in the two sets of
variables.

We start by noting that, at zeroth order in the Weyl background, the
new variables are equal to the original variables $S$, $T^a$ and
$Y^0$, so that it is convenient to write
\begin{eqnarray}
  \label{eq:new-variables}
  {\tilde Y}^0 &=& Y^0 + \Delta Y^0 \;,\nonumber\\
  {\tilde S} &=& S + \Delta S \;,\nonumber\\
  {\tilde T}^a &=& T^a + \Delta T^a \;.
\end{eqnarray}
Here the changes $\Delta$ are induced by the Weyl background, which
is encoded in the dependence on the field $\Upsilon$.  Observe that
the changes $\Delta$ will be non-holomorphic due to the reality
property of the map (2.13), thus leading to a change of complex
structure. This non-holomorphicity is thus not related to the fact
that the function $\Omega$ is not necessarily harmonic. Using
\eqref{eq:tilde-fields-Y} and the explicit expressions \eqref{eq:F_I}
we derive the following six real equations for these changes,
\begin{eqnarray}
  \label{eq:changes-1a}
  \mathrm{Re}\left[\Delta Y^0\right] &=& 0\;, \\[4mm]
    \label{eq:changes-1b}
  \mathrm{Im}\left[Y^0\,\Delta S + S\,\Delta Y^0  + \Delta S\,\Delta
  Y^0 \right]   &=& 0\;, \\[4mm]  
    \label{eq:changes-1c}
  \mathrm{Im}\left[Y^0\,\Delta T^a + T^a\,\Delta Y^0 + \Delta T^a\,\Delta
  Y^0\right] &=& 0\;, 
\end{eqnarray}
and 
\begin{eqnarray}
  \label{eq:changes-2a}
  \mathrm{Re}\left[(Y^0+\Delta Y^0) (T+\Delta T)^a\eta_{ab}(T+\Delta 
  T)^b - Y^0\,T^a\eta_{ab}T^b \right] =\qquad && \nonumber \\[2mm]
   \frac{1}{Y^0} \,\frac{\partial\Omega}{\partial S} + 
   \frac1{\bar Y^0} \,\frac{\partial\Omega}{\partial\bar S}\;,&&
   \\[4mm] 
     \label{eq:changes-2b}
  \mathrm{Re}\left[(Y^0+\Delta Y^0)(S+\Delta S)\,(T+\Delta 
  T)^a - Y^0 S\,T^a \right] =\qquad &&\nonumber \\[2mm]
   \frac{1}{2\,Y^0}
   \,\frac{\partial\Omega}{\partial T_a} +
   \frac{1}{2\,\bar Y^0} \,\frac{\partial\Omega}{\partial\bar
  T_a}\;,&& \\[4mm]
  \label{eq:changes-2c}
  \mathrm{Im}\left[(Y^0+\Delta Y^0)(S+\Delta S) (T+\Delta
  T)^a\eta_{ab}(T+\Delta T)^b - Y^0 S\, T^a\eta_{ab}T^b \right]
  =\qquad &&
   \nonumber \\[2mm]
   - \mathrm{i}\,\left[- \frac{\partial\Omega}{\partial Y^0} +
  \frac{\partial\Omega}{\partial\bar Y^0} +
  \frac{S}{Y^0}\,\frac{\partial\Omega}{\partial S}-
  \frac{\bar S}{\bar Y^0}\,\frac{\partial\Omega}{\partial\bar S}+
  \frac{T^a}{Y^0}\,\frac{\partial\Omega}{\partial T^a} -
  \frac{\bar T^a}{\bar Y^0}\,\frac{\partial\Omega}{\partial\bar
  T^a}\right]  \;,
\end{eqnarray}
where here and henceforth indices $a,b,\ldots$ are lowered and raised
with $\eta_{ab}$ and its inverse $\eta^{ab}$. The left-hand side of
these equations are polynomials of at most fourth degree in $\Delta
S$, $\Delta T^a$, $\Delta Y^0$, and their complex conjugates. The
equations \eqref{eq:changes-1b} and \eqref{eq:changes-1c} can
conveniently be written as
\begin{eqnarray}
  \label{eq:Delta-ST-im}
  (Y^0+ \Delta Y^0)\,\Delta S - (\bar Y^0- \Delta Y^0)\,\Delta\bar S &=& - 
  (S+\bar S) \,\Delta Y^0\,,\nonumber\\
    (Y^0+ \Delta Y^0)\,\Delta T^a - (\bar Y^0- \Delta Y^0)\,\Delta\bar T^a 
  &=& - (T+\bar T)^a \,\Delta Y^0\,.
\end{eqnarray}
The six equations can be solved by iteration which will lead to
explicit power expansions in first-order derivatives of $\Omega$.

Before proceeding we note that it is convenient to write all the
$\Omega$-dependent terms in the form of one real and one complex
combination,
\begin{eqnarray}
  \label{eq:Omega-combinations}
  \mathcal{T}_a&=& \frac{1}{Y^0}
   \,\frac{\partial\Omega}{\partial T^a} +
   \frac{1}{\bar Y^0} \,\frac{\partial\Omega}{\partial\bar T^a} \;,
 \nonumber\\[.2ex] 
  \mathcal{U}&=&   \frac{\partial  \Omega}{\partial Y^0} 
  -\frac{\partial \Omega}{\partial {\bar Y}^0} - \frac{(S + \bar S)}{{Y}^0}\,  
  \frac{\partial \Omega}{\partial S} 
  + \frac{(T + \bar T)^a}{{\bar Y}^0} \,
  \frac{\partial \Omega}{\partial \bar T^a}  \;.  
\end{eqnarray}
With these definitions, we first consider \eqref{eq:changes-2b},
which, with the help of \eqref{eq:Delta-ST-im}, takes the form, 
\begin{equation}
  \label{eq:changes-2b-2}
  (S + \bar S +\Delta S+\Delta \bar S)(Y^0+\Delta Y^0) \,\Delta T^a
  +(T+\bar T)^a \,\bar Y^0 \,\Delta \bar S = \mathcal{T}^a  \;. 
\end{equation}
Likewise \eqref{eq:changes-2a} can be written, again with the help of
\eqref{eq:Delta-ST-im}, as 
\begin{eqnarray}
  \label{eq:changes-2a-2-alt}
  &&
  (Y^0 + \Delta Y^0)\Delta T_a(2T+2\bar T + \Delta T+ \Delta\bar T)^a =
  \nonumber \\
  &&
   -\Delta Y^0 (T+\bar T)_a (T+\bar T +\Delta \bar T )^a   
   - 2\,(S+\bar S)^{-1} \big[\mathcal{U}+\bar{\mathcal{U}}-(T+\bar T)^a
   \mathcal{T}_a \big]   \;.  
\end{eqnarray}
Finally, \eqref{eq:changes-2c} can be written in the form, 
\begin{eqnarray}
  \label{eq:2c-altern}
  &&
  (Y^0+\Delta Y^0) \Delta S(T^2-\bar T^2)  \nonumber\\
  &&  + (\bar Y^0-\Delta Y^0)\,\Delta \bar T^a\left[(2T+\Delta T)_a
  (S+\Delta S) - (2\bar T+\Delta \bar T)_a (\bar S+\Delta \bar
  S)\right ] \nonumber\\  
  && 
  -\Delta Y^0 \,(T+\bar T)^a \left[ S (T+\bar T+\Delta T)_a 
  + \Delta S(2\,T+ \Delta T)_a \right] = \nonumber\\
    && \quad
    2\,(S+\bar S)^{-1} \big[-S\,\mathcal{U}
  + \bar S\,\bar{\mathcal{U}}  + (S\,T^a-\bar S\,\bar T^a)
  \mathcal{T}_a\big]  \;.
\end{eqnarray}

By taking suitable linear combinations we now write these equations in
a form such that the first-order term is just proportional to either
$\Delta S$, $\Delta T^a$, or $\Delta Y^0$. This will enable us to
directly obtain the first-order results, whereas the higher-order ones
will follow from iteration. The equation for $\Delta S$ is as follows,
\begin{eqnarray}
  \label{eq:Delta-S-equation} 
  && Y^0 \left(T + {\bar T} \right)^2 \, \Delta S =  2\,
    \bar{\mathcal{U}} + \Delta Y^0 \, \Delta S\, 
    (T+\bar T)^a \, (T + \bar T + \Delta T)_a  \nonumber\\[.2ex] 
    &&\quad
    -(\bar Y^0-\Delta Y^0) \,\Delta \bar T^a\left[\Delta
    S\,(2\,T+2\,\bar T +\Delta T)_a -(S +\bar S+ \Delta \bar
    S)\Delta\bar T_a \right]  \,. 
 \end{eqnarray}
For $\Delta T^a$, the expression takes the form,
\begin{eqnarray}
  \label{eq:Delta-T-equation}
  &&
  Y^0(S+\bar S)(T+\bar T)^2\,\Delta T^a = (T+\bar T)^2 \,\mathcal{T}^a
  - 2\,(T+\bar T)^a \,\mathcal{U} \nonumber \\[.2ex] 
  &&
  \quad
  + (T+\bar T)^a\, \Delta Y^0\,\Delta\bar S \,(T+\bar T)^b(T+\bar
  T+\Delta\bar T)_b \nonumber\\[.2ex] 
  &&
  \quad
  + (T+\bar T)^a(Y^0+ \Delta Y^0)\,\Delta T^b\left[\Delta\bar
  S(2\,T+2\,\bar T+ \Delta\bar T)_b -(S+\bar S + \Delta S)\Delta
  T_b\right] \nonumber\\[.2ex] 
  &&
  \quad
  -\Delta T^a(T+\bar T)^2 \left[(Y^0+\Delta Y^0)(\Delta S+\Delta\bar
  S) + \Delta Y^0(S+\bar S)\right] \,,
\end{eqnarray}
and for $\Delta Y^0$, one obtains, 
\begin{eqnarray}
  \label{eq:Delta-Y0-equation}
  &&
  (S+\bar S)(T+\bar T)^2 \Delta Y^0 =  2\,
  (\mathcal{U}-\bar{\mathcal{U}}) \nonumber \\[.2ex] 
    &&\quad
    - \Delta Y^0 \left[2(\Delta S+ \Delta\bar S)\,
    (T+\bar T)^2 + (T + \bar T)^a(\Delta T_a \Delta S +\Delta\bar T_a
  \Delta \bar S)\right]
  \nonumber\\[.2ex]  
    &&\quad
    +(\bar Y^0-\Delta Y^0) \,\Delta \bar T^a\left[\Delta
    S\,(2\,T+2\,\bar T+ \Delta T)_a -(S +\bar S+ \Delta \bar
    S)\Delta\bar T_a \right]  \nonumber\\[.2ex] 
    &&\quad
    -(Y^0+\Delta Y^0) \,\Delta T^a\left[\Delta\bar S\,(2\,T+2\,\bar T+
  \Delta\bar T)_a
  -(S +\bar S+ \Delta S)\Delta T_a \right]  \,.   
\end{eqnarray}

These three equations constitute quartic polynomials in $\Delta S$,
$\Delta T^a$ and $\Delta Y^0$, which can be can, in principle, be
solved by iteration. Obviously, the full solution for $\Delta S$,
$\Delta T^a$ and $\Delta Y^0$ will then take the form of an infinite
series of products of the functions $\mathcal{U}$ and $\mathcal{T}_a$.
To simplify the iteration to higher orders, it is convenient to use
\eqref{eq:changes-2b-2} and \eqref{eq:changes-2a-2-alt} once more for
the higher-order terms of
\eqref{eq:Delta-S-equation}-\eqref{eq:Delta-Y0-equation}. We find,
respectively,
\begin{eqnarray}
  \label{eq:Delta-S-equation-2} 
  Y^0 \left(T + {\bar T} \right)^2 \, \Delta S &=&  2\,\bigg\{
    \bar{\mathcal{U}} 
  + \frac{\Delta S \,\left[\mathcal{U} +\bar{\mathcal{U}}
    - (T+\bar T)^a \mathcal{T}_a\right]}{S+\bar S} \bigg\}
    \nonumber\\[.2ex] 
    &&{}
    + \Delta \bar T^a\left[\mathcal{T}_a - (T+\bar T)_a Y^0 \Delta S\right]
    \,,\\[1.4ex]
  \label{eq:Delta-T-equation-2}  
  Y^0(S+\bar S)(T+\bar T)^2\,\Delta T^a &=& (T+\bar T)^2
  \left[\mathcal{T}^a 
  -\Delta T^a\, (Y^0+\bar Y^0) \Delta\bar S \right]   \nonumber \\[.2ex] 
  &&{}
  - 2\,(T+\bar T)^a \,\bigg\{ \mathcal{U}
  + \frac{\Delta \bar S \,\left[\mathcal{U} +\bar{\mathcal{U}} -
  (T+\bar T)^b \mathcal{T}_b\right]} {S+\bar S} \bigg\}
  \nonumber\\[.2ex]  
  &&{}
  - (T+\bar T)^a\,\Delta T^b\left[\mathcal{T}_b -(T+\bar T)_b \bar Y^0
  \Delta\bar S \right] \,, \\[1.4ex]
  \label{eq:Delta-Y0-equation-2}
  (S+\bar S)(T+\bar T)^2 \Delta Y^0 &=&
  2\,\bigg\{\mathcal{U}-\bar{\mathcal{U}}  
    + \frac{(\Delta \bar S-\Delta S)\,\left[\mathcal{U} +\bar{\mathcal{U}}
    - (T+\bar T)^a \mathcal{T}_a\right]}{S+\bar S}\bigg\}   \nonumber\\[.2ex]  
    &&{} 
    + (\Delta T-\Delta\bar T)^a \mathcal{T}_a 
    - (T+\bar T)^2\,\Delta Y^0 (\Delta S+ \Delta\bar S)
     \nonumber\\[.8ex]
    &&{}
    -(T+\bar T)^a 
    \left[\bar Y^0 \Delta T_a \Delta\bar S - Y^0 \Delta \bar T_a
      \Delta S\right]\,.    
\end{eqnarray}
The lowest-order solution can be read off from 
\eqref{eq:Delta-S-equation-2}~-~\eqref{eq:Delta-Y0-equation-2}, and
takes the form,
\begin{eqnarray}
  \label{eq:first-order-soln}
  Y^0 \left(T + {\bar T} \right)^2 \, \Delta S &=&  2\,
  \bar{\mathcal{U}}\,,  \nonumber \\[.2ex] 
  Y^0(S+\bar S)(T+\bar T)^2\,\Delta T^a &=& (T+\bar T)^2 \,\mathcal{T}^a
  - 2\,(T+\bar T)^a \,\mathcal{U}\,,
  \nonumber \\[.2ex] 
  (S+\bar S)(T+\bar T)^2 \Delta Y^0 &=&  2\,
  (\mathcal{U}-\bar{\mathcal{U}}) \,. 
\end{eqnarray}
Resubstituting this result on the right-hand side of
\eqref{eq:Delta-S-equation-2}~-~\eqref{eq:Delta-Y0-equation-2}, yields
the results to second order,  
\begin{eqnarray}
  \label{eq:Delta-S-second-order}
   Y^0 \left(T + {\bar T} \right)^2 \, \Delta S &\approx&
   2\,\bar{\mathcal{U}} 
  + \frac{4\,\bar{\mathcal{U}}\left[\mathcal{U} +\bar{\mathcal{U}} 
    - (T+\bar T)^a \mathcal{T}_a\right]}{Y^0(S+\bar S)(T+\bar T)^2} 
    \nonumber\\[.4ex] 
    &&{}
    + \frac{1}{\bar Y^0 (S+\bar S)} \,
\left[\mathcal{T}^a - \frac{2\,(T+\bar T)^a \, 
   \bar{\mathcal{U}}}{ (T+\bar T)^2} \right]^2
    \,,\\[1.4ex]
    \label{eq:Delta-T-second-order}  
    Y^0(S+\bar S)(T+\bar T)^2\,\Delta T^a &\approx& (T+\bar T)^2
    \,\mathcal{T}^a  -  2\,(T+\bar T)^a \,\mathcal{U} 
  \nonumber\\[.4ex]   
  &&{}
  -\frac{2\, (Y^0+\bar Y^0) \,\mathcal{U}
   \left[(T+\bar T)^2 \,\mathcal{T}^a
  - 2\,(T+\bar T)^a \,{\mathcal{U}}\right]}{\vert Y^0\vert^2(S+\bar
  S)(T+\bar T)^2}    \nonumber \\[.2ex]  
  &&{}
  - 4\,(T+\bar T)^a \, \frac{\mathcal{U}\, \left[\mathcal{U}
  +\bar{\mathcal{U}} - 
  (T+\bar T)^b \mathcal{T}_b\right]} {\bar Y^0(S+\bar S)(T+\bar T)^2}
  \nonumber\\[.2ex]  
  &&{}
  - \frac{(T+\bar T)^a}{Y^0 (S+\bar S)} \, \left[\mathcal{T}^b -
  \frac{2(T+\bar T)^b\,\mathcal{U}}{(T+\bar T)^2} \right]^2 \,, \\[1.4ex]
  \label{eq:Delta-Y0-second-order}
  (S+\bar S)(T+\bar T)^2 \Delta Y^0 &\approx&
  2\,\mathcal{U}  
    + \frac{4\,\mathcal{U}\left[2\,\bar{\mathcal{U}}
    - (T+\bar T)^a \mathcal{T}_a\right]}{\bar Y^0 (S+\bar S)(T+\bar
  T)^2 }    \nonumber\\[.2ex]  
    &&{}
    +  \frac{1}{Y^0 (S+\bar S)} \left[\mathcal{T}^a - \frac{2\,(T+\bar
  T)^a\,{\mathcal{U}} }{(T+\bar T)^2} \right]^2
%
     \nonumber\\[.6ex]
    &&{}
    - \mathrm{h.c.} \,. 
\end{eqnarray}

At the end of this section we briefly return to the transformation
rules under S- and T-duality of the new variables. Obviously the
previous equations should be consistent with these duality
transformations. To verify this one first determines the
transformation rules of $\Delta S$, $\Delta T^a$ and $\Delta Y^0$,
which follow straightforwardly from their definition
\eqref{eq:new-variables} and the transformations acting on the old and
the new fields. In this way one obtains the following results under
S-duality, 
\begin{eqnarray}
  \label{eq:s-dual-Delta}
  \Delta S &\to& \frac{\Delta S}{\Delta_\mathrm{S} (\Delta_\mathrm{S}
  + \mathrm{i} c\,\Delta S) } \;, \nonumber \\[.2ex]
   \Delta T^a &\to & \Delta T^a -\frac{\mathrm{i}\,c}{\Delta_\mathrm{S}
  (Y^0)^2} \,\frac{\partial\Omega}{\partial T_a}\;, \nonumber\\[.4ex]
   \Delta Y^0&\to& \tfrac12(\Delta_\mathrm{S}+\bar \Delta_\mathrm{S})
  \Delta Y^0 + 
  \tfrac12\mathrm{i}\,c\left[\Delta S(Y^0+\Delta Y^0) + \Delta\bar
  S(\bar Y^0-\Delta Y^0) \right]\;. 
\end{eqnarray}
Likewise, under T-duality one finds,
\begin{eqnarray}
  \label{eq:s-dual-Delta}
  \Delta S &\to& \Delta S +\frac{2}{\Delta_\mathrm{T}(Y^0)^2} \left[Y^0
  \frac {\partial\Omega}{\partial Y^0}
  - T^a\frac{\partial\Omega}{\partial T^a}\right] \;, \nonumber \\[.2ex]
   \Delta T^a &\to & \frac{\Delta T^a}{(T+\Delta T)^2} +
  \frac{T^a}{(T+\Delta T)^2 \Delta_\mathrm{T}} 
  \,\left[-2\,T^b \Delta T_b - (\Delta T)^2 
  +\frac{2}{(Y^0)^2} \,\frac{\partial\Omega}{\partial S}\right] \;,
  \nonumber\\[.6ex] 
   \Delta Y^0&\to& 
  \tfrac12 (T^2 +\bar T^2) \,\Delta Y^0 \nonumber\\
  &&{} 
  + \tfrac12\left[(Y^0+\Delta Y^0) \Delta T^a(2\,T+\Delta 
   T)_a -(\bar Y^0-\Delta Y^0) \Delta\bar T^a(2\,\bar T+ \Delta\bar
   T)_a  \right] 
   \nonumber \\
   &&{}
     -\left[\frac{1}{Y^0} \frac{\partial\Omega}{\partial S}
    -\frac{1}{\bar Y^0} \frac{\partial\Omega}{\partial\bar S} \right] \;, 
\end{eqnarray}
where, in the T-duality transformation of $\Delta Y^0$, we made use of
\eqref{eq:changes-2a} in order to write the right-hand side in a form
that is manifestly imaginary.

To verify the consistency we also need the transformations of the
functions $\mathcal{U}$ and $\mathcal{T}^a$ under S- and T-duality,
which follow from the results listed in section
\ref{sec:S-T-dualities}. Under S-duality these transformations take
the following form, 
\begin{eqnarray}
  \label{eq:T-U-Sdual}
  \mathcal{T}_a&\to& \frac{1}{\Delta_\mathrm{S} Y^0}
  \,\frac{\partial\Omega}{\partial T^a} + \frac{1}{\bar
  \Delta_\mathrm{S} \bar Y^0} 
  \,\frac{\partial\Omega}{\partial \bar T^a} \,,\nonumber\\
  \mathcal{U}&\to& \frac{\mathcal{U}}{\bar\Delta_\mathrm{S}}
  +\frac{\mathrm{i}c}{\vert\Delta_\mathrm{S}\vert^2 (Y^0)^2} \,
  \mathcal{T}^a  \frac{\partial\Omega}{\partial T^a} 
  +\frac{c^2(S+\bar S)}{2\,\vert\Delta_\mathrm{S}\vert^2\,
  \Delta_\mathrm{S}\, (Y^0)^3} \, 
   \frac{\partial\Omega}{\partial T^a}\eta^{ab}
  \frac{\partial\Omega}{\partial T^b}  \,.  
\end{eqnarray}
Under T-duality one derives the following transformations, 
\begin{eqnarray}
  \label{eq:T-U-Tdual}
    \mathcal{T}^a&\to& \mathcal{T}^a + \frac{2\,T^a}{\Delta_\mathrm{T}}
  \,\left(\frac{\partial\Omega}{\partial Y^0} -\frac{T^b}{Y^0}
  \,\frac{\partial\Omega}{\partial T^b}\right) +
  \frac{2\,\bar T^a}{\bar\Delta_\mathrm{T}}
  \,\left(\frac{\partial\Omega}{\partial \bar Y^0} -\frac{\bar
  T^b}{\bar Y^0}
  \,\frac{\partial\Omega}{\partial\bar T^b}\right)\,,\nonumber\\ 
  \mathcal{U} &\to& 
  \frac{\mathcal{U}}{\Delta_\mathrm{T}} 
  + \frac{(T+\bar T)^2} {\vert\Delta_\mathrm{T}\vert^2}
  \left(\frac{\partial\Omega}{\partial \bar Y^0} 
     -\frac{\bar T^a} {\bar Y^0}  
  \,\frac{\partial\Omega}{\partial\bar T^a}\right) \nonumber \\
   &&{}
   -\frac{2}{(\Delta_\mathrm{T})^2\, (Y^0)^2}
  \frac{\partial\Omega}{\partial S} 
\left(\frac{\partial\Omega}{\partial Y^0} 
     -\frac{T^a} {Y^0}  
  \,\frac{\partial\Omega}{\partial T^a}\right) 
  \nonumber\\ 
  &&{}
  +\frac{2}{\vert\Delta_\mathrm{T}\vert^2}\left( 
  \frac1{(Y^0)^2} \frac{\partial\Omega}{\partial S} + 
  \frac1{\vert Y^0\vert^2} \frac{\partial\Omega}{\partial S} +
  \frac1{(\bar Y^0)^2} \frac{\partial\Omega}{\partial\bar S} \right)
  \left(\frac{\partial\Omega}{\partial \bar Y^0} 
    -\frac{\bar T^a}{\bar Y^0}   
  \,\frac{\partial\Omega}{\partial\bar T^a}\right) \;.  
\end{eqnarray}
Here $\Delta_\mathrm{S}$ and $\Delta_\mathrm{T}$ were defined in
\eqref{eq:Delta-S} and \eqref{eq:Delta-T}.  With the above results
\eqref{eq:s-dual-Delta} - \eqref{eq:T-U-Tdual} one can verify that the
equations \eqref{eq:Delta-S-equation-2} -
\eqref{eq:Delta-Y0-equation-2} for $\Delta S$, $\Delta T^a$ and $\Delta
Y^0$ are fully consistent with S- and T-duality.

\section{Heterotic $N=4$ supersymmetric string compactifications}
\label{sec:heterotic-example}
\setcounter{equation}{0}
As a first application we demonstrate the results of the previous
section in a specific example, which is relevant in the context of
$N=4$ supersymmetric models. Namely we assume that $\Omega$ depends
only on $S$ and $\bar S$. Homogeneity then implies that $\Omega$ will
depend linearly on $\Upsilon$ and its complex conjugate. In this case
we have
\begin{equation}
  \label{eq:TU-example}
  \mathcal{U} = -\frac{S+\bar S}{Y^0} \,\frac{\partial\Omega}{\partial
    S}\;,   \qquad\mathcal{T}_a=0\,.
\end{equation}
By direct inspection it follows that the S-duality transformations
take the simple form given by (\ref{eq:ST-Sdual}) and that the
equations (\ref{eq:S-invariance}) are satisfied provided that $\Omega$
is an S-duality invariant function. For T-duality the situation is more
subtle. The equations (\ref{eq:T-invariance}) are manifestly
satisfied, but the transformation rules under T-duality take a
complicated form,
\begin{equation}
  \label{eq:T-dual-example}
  S\to S\;, \qquad T^a\to \frac{T^a}{\Delta_\mathrm{T}} \;, \qquad
  Y^0\to \Delta_\mathrm{T} \,Y^0\;,
\end{equation}
where $\Delta_\mathrm{T}$ is still given by (\ref{eq:Delta-T}). Hence
this example is consistent with both dualities. Under S- and
T-duality $\mathcal{U}$ now transforms according to
\begin{equation}
  \label{eq:U-ST}
  \mathcal{U}\to \frac{\mathcal{U}}{\bar\Delta_\mathrm{S}} \,,\qquad
  \mathcal{U}\to \frac{\mathcal{U}}{\Delta_\mathrm{T}} \,.
\end{equation}
respectively. 

As it turns out, it is convenient to introduce an S-duality invariant
variable $\mathcal{V}$, defined by
\begin{equation}
  \label{eq:def-cal-V}
  \mathcal{V} = -\frac{\mathcal{U}}{Y^0(S+\bar S)(T+\bar T)^2} =
  \frac1{(Y^0)^2\,(T+\bar T)^2}\,\frac{\partial\Omega}{\partial S}
  \;,
\end{equation}
so that the definition (\ref{eq:Delta-T}) reads
\begin{equation}
  \label{eq:Delta-T-new}
  \Delta_\mathrm{T}= T^2 +  2\,(T+\bar T)^2\,\mathcal{V} \,. 
\end{equation}
Under T-duality $\mathcal{V}$ transforms non-trivially according to
\begin{eqnarray}
  \label{eq:cal=V-T}
  \mathcal{V}\to \frac{\bar\Delta_\mathrm{T}}{\Delta_\mathrm{T}}\, 
  \mathcal{V}\, 
  \left[ 1 + 2(\mathcal{V}+\bar{\mathcal{V}})- 2
    \left(\frac{\bar \Delta_\mathrm{T}}{\Delta_\mathrm{T}} \,
      \mathcal{V}+ \frac{\Delta_\mathrm{T}}{\bar\Delta_\mathrm{T}}
      \,\bar{\mathcal{V}}\right)\right]^{-1}  \,.
\end{eqnarray}

To solve the various equations of the previous section, we note that
$\Delta T^a$ must be proportional to $(T+\bar T)^a$ in this
example. The proportionality factor turns out to be a function of the
variable $\mathcal{V}$ and its complex conjugate, and is therefore
S-duality invariant. The full expressions for $\Delta T^a$, $\Delta S$ 
and $\Delta Y^{0}$ then take the form,
\begin{eqnarray}
  \label{eq:Delta-f}
    \Delta T^a&=& f\,(T+\bar T)^a \,,\nonumber \\[.4ex]
    \Delta S&=& -\,\frac{\bar f\,(S+\bar S) \,\bar Y^0} {(1+\bar f) Y^0+
    \bar f\,\bar Y^0} \,, \nonumber  \\[.2ex]
    \Delta Y^0&=& \frac{\bar f\,\bar Y^0 -f\, Y^0}{1+ f+\bar f}\,,  
\end{eqnarray}
or, alternatively,
\begin{eqnarray}
  \label{eq:f-tilde}
  \tilde T^a&=& (1+f)\,T^a +f\, \bar T^a \,,\nonumber \\[.4ex]
  \tilde S&=& \frac{(1+\bar f)\,S\,Y^0 - \bar f\,\bar S\,\bar Y^0}
  {(1+\bar f) Y^0+ \bar f\,\bar Y^0} \,, \nonumber  \\[.2ex]
  \tilde Y^0&=& \frac{(1+\bar f)Y^0 +\bar f\,\bar Y^0}{1+f+\bar f}\,, 
\end{eqnarray}
where the function $f(\mathcal{V},\bar{\mathcal{V}})$ is defined by a
quadratic equation, 
\begin{equation}
  \label{eq:eq-f}
  \vert f\vert^2+ f = 2\, \mathcal{V}\,.
\end{equation}
From this result it follows that 
\begin{equation}
  \label{eq:r-1}
  (1+f+\bar f)^2 = 1 +4(\mathcal{V} +\bar{\mathcal{V}}) +4(\mathcal{V}
  -\bar{\mathcal{V}})^2 \geq 0 \,. 
\end{equation}
The solution for $f(\mathcal{V},\bar{\mathcal{V}})$ reads as follows, 
\begin{equation}
   \label{eq:f-sol}
   f= -\tfrac12 +\mathcal{V} -\bar{\mathcal{V}} \pm 
   \tfrac12\sqrt{1 +4(\mathcal{V} +\bar{\mathcal{V}}) +4(\mathcal{V}
  -\bar{\mathcal{V}})^2}\,,
\end{equation}
where one must adopt the plus sign in order to correctly reproduce the
situation where $\Omega$ vanishes. 

The new fields (\ref{eq:f-tilde}) transform indeed as required. With
regard to T-duality the following transformation of $f$ under
T-duality,
\begin{equation}
  \label{eq:f-T-duality}
  f\to \frac{\bar\Delta_\mathrm{T}}{\Delta_\mathrm{T}}\, f \,\left[
    1+f -\frac{\bar\Delta_\mathrm{T}}{\Delta_\mathrm{T}}\,f
  \right]^{-1} \;, 
\end{equation}
is sufficient to ensure the correct T-duality transformations for the
fields (\ref{eq:f-tilde}). Incidentally, (\ref{eq:f-T-duality})
can be rewritten as
\begin{equation}
  \label{eq:f-T-duality-2}
   f\to \frac{\bar\Delta_\mathrm{T}} {\tilde T^2}\;f \,, 
\end{equation}
by making use of
\begin{eqnarray}
  \label{eq:new-T-square}
  \tilde T^2 &=& (1+f) T^2 -f\,\bar T^2 +f(1+f)(T+\bar T)^2 \nonumber\\
   &=& (1+ f)\Delta_\mathrm{T} -f\,\bar\Delta_\mathrm{T}\,.
\end{eqnarray}

Subsequently we evaluate the first term of the Hesse potential
\eqref{eq:GenHesseP2},
\begin{eqnarray}
  \label{eq:hesse-1-exact}
    -\mathrm{i} (\bar Y^I F_I-Y^I\bar F_I)&=&
    -\vert \tilde Y^0\vert^2 (\tilde S+\bar{\tilde S})(\tilde
    T+\bar{\tilde T})^2  \left[1 + \frac{2\,\vert f\vert^2} {1+f+\bar
        f}  \right] \nonumber\\[2mm]
    &=&
    -\vert \tilde Y^0\vert^2 (\tilde S+\bar{\tilde S})(\tilde
    T+\bar{\tilde T})^2  \frac{ 1 + 2\,(\mathcal{V}+\bar{\mathcal{V}})}
     {\sqrt{1 +4(\mathcal{V} +\bar{\mathcal{V}}) +4(\mathcal{V}
         -\bar{\mathcal{V}})^2}} \;,
\end{eqnarray}
where the sign adopted in the last expression is consistent with the
sign choice noted below (\ref{eq:f-sol}). To evaluate this result we
used the following equations,
\begin{eqnarray}
  \label{eq:Delta-f-real}
  Y^0&=& (1+f)\tilde Y^0 -\bar f\,\bar{\tilde Y}^0\,, \nonumber \\[1mm]
  (\tilde T+\bar{\tilde T})^a&=& (1+f+\bar f) (T+\bar T)^a
  \,,\nonumber   \\ [2mm]
  \tilde S+\bar{\tilde S} &=& \frac{(1+f+\bar f)\,\vert Y^0\vert^2\,
    (S+\bar S)} {\vert(1+\bar f)\,Y^0 +\bar 
    f\,\bar Y^0\vert^2 } \,.
\end{eqnarray}
Subsequently it is convenient to introduce a quantity $\lambda$, 
\begin{eqnarray}
  \label{eq:def-lambda}
  \lambda &=&{} \frac{\bar f}{1+f}\;\frac{\bar{\tilde Y}^0}{\tilde
    Y^0}\nonumber\\ 
  &=&{} - \frac{1 +2(\mathcal{V}-\bar{\mathcal{V}}) -\sqrt{1
        +4(\mathcal{V} +\bar{\mathcal{V}}) +4(\mathcal{V}
        -\bar{\mathcal{V}})^2}} 
    {1 +2(\mathcal{V}-\bar{\mathcal{V}}) +\sqrt{1 +4(\mathcal{V}
          +\bar{\mathcal{V}}) +4(\mathcal{V} 
         -\bar{\mathcal{V}})^2}}\;\;\frac{\bar{\tilde Y}^0}{\tilde
    Y^0} \,,
\end{eqnarray}
so that 
\begin{equation}
  \label{eq:new-Delta-S}
  \Delta S=  - (\tilde S+\bar{\tilde S})\, \frac{\lambda}{1-\lambda}
  \,. 
\end{equation}
The newly defined quantity $\lambda$ is invariant under
T-duality, while under S-duality it transforms with a phase factor,
\begin{equation}
  \label{eq:s-var-lambda}
  \lambda\to  \frac{\bar{\tilde\Delta}_\mathrm{S}}
  {\tilde\Delta_\mathrm{S}} \;\lambda \,,
\end{equation}
where $\tilde\Delta_\mathrm{S}= d+ \mathrm{i}c\,\tilde S$.

We thus obtain the following expression for $\mathcal{V}$,
\begin{equation}
  \label{eq:cal-V-cont}
  \mathcal{V} = \left(
  \frac{1}{\tilde Y^0} +\frac{\lambda}{\bar{\tilde Y}^0}\right)^2\; 
  \frac{1}{(\tilde T+\bar{\tilde T})^2} \;
  \frac1{(1-\lambda)^2}\;\frac{\partial\Omega}{\partial S} \,.  
\end{equation}
It remains to write the derivative of $\Omega(S, \bar S)$ as a
function of the new field $\tilde S$ and its complex conjugate. This
can simply be done by writing $\Omega$ as 
\begin{equation}
  \label{eq:cal-V-cont-2}
  \Omega(S,\bar S) =  
  \Omega\Big(\tilde S +
  \frac{(\tilde S+\bar{\tilde S})\lambda}{1-\lambda}, \bar{\tilde S} +
  \frac{(\tilde S+\bar{\tilde S})\bar\lambda}{1-\bar\lambda}\Big)\;,
\end{equation}
and Taylor-expanding $(1-\lambda)^{-2}\,\partial_S\Omega$ in
$\lambda$ and $\bar\lambda$. This leads to a double expansion in
multiple covariant derivatives, 
\begin{equation}
  \label{eq:expand-tildeS}
  \frac{\partial_S\Omega(S,\bar S)}{(1-\lambda)^2} = \sum_{m=1.n=0}^\infty
  \;c_{(m,n)}(\vert\lambda\vert) \; \lambda^{m-1}\bar \lambda^n\,
  (\tilde S+\bar{\tilde S})^{m+n-1}\, \;
  (D_{\tilde S})^m \;(\bar D_{\tilde{\bar S}})^n\, \Omega(\tilde
  S,\tilde{\bar S}) \,, 
\end{equation}
with respect to the new fields. Here the $c_{(m,n)}(\lambda)$ are
functions of $\vert\lambda\vert$. The covariant derivatives are
defined as follows. A modular form $\omega_{p,q}(S,\bar S)$
of degree $(p,q)$ transforms according to $\omega_{p,q}(S,\bar
S)\to [\Delta(S)]^p\, [\bar\Delta(\bar S)]^q\,\omega_{p,q}(S,\bar
S)$. Its covariant derivative $D_S$ is then defined by
\begin{eqnarray}
  \label{eq:S-cov-derivative}
  D_S \,\omega_{p,q}(S,\bar S) = \left(\partial_S + \frac{p}{S+\bar S}
    \right)\omega_{p,q}(S,\bar S)\,, 
\end{eqnarray}
and transforms as $D_S \,\omega_{p,q}(S,\bar S)\to [\Delta(S)]^{p+2}\,
[\bar\Delta(\bar S)]^q\,D_S\,\omega_{p,q}(S,\bar S)$. The covariant
derivative with respect to $\bar S$ is defined likewise. 

To obtain the Hesse potential \eqref{eq:GenHesseP2}, it remains to add
$4\,\Omega$ to
(\ref{eq:hesse-1-exact}). Using the same strategy as above, we can
derive an expression similar to \eqref{eq:expand-tildeS} for $\Omega$
expressed in $\tilde S$ and its complex conjugate,
\begin{equation}
  \label{eq:expand-tildeS-Omega}
  \Omega(S,\bar S) = \sum_{m,n=0}^\infty
  \;d_{(m,n)}(\vert\lambda\vert) \; \lambda^{m}\bar \lambda^n\,
  (\tilde S+\bar{\tilde S})^{m+n}\, \;
  (D_{\tilde S})^m \;(\bar D_{\tilde{\bar S}})^n\, \Omega(\tilde
  S,\tilde{\bar S}) \,. 
\end{equation}

As an illustration we have evaluated all contributions up to third
order in $\Upsilon, \bar \Upsilon$. The first term in the Hesse
potential \eqref{eq:GenHesseP2}, which in the case at hand is given by
(\ref{eq:hesse-1-exact}), can be expanded in powers of $\mathcal{V}$
and $\lambda$, making use of (\ref{eq:def-lambda}) and
(\ref{eq:cal-V-cont}). In this way the result is expressed in terms of
the new fields $\tilde Y^0$, $\tilde S$ and $\tilde T^a$. Up to third
order one obtains (where on the right-hand side we have 
suppressed the tilde for clarity of notation),
\begin{eqnarray}
  \label{eq:hesse-1-ext}
    -\mathrm{i} (\bar Y^I F_I-Y^I\bar F_I)&\approx&{}
    -\vert Y^0\vert^2 ( S+\bar{S})(T+\bar{T})^2 
    \nonumber\\
    && {}
    - \frac{8 \, (S+\bar{S})\,
      \big\vert\partial_S\Omega\big\vert^2} {|{Y}^0|^2
      (T+\bar{T})^2} \left(1 +
      \frac{4\,(S+\bar{S})\, \partial_S \partial_{\bar S}\Omega}
      {\vert{Y}^0|^2 (T+\bar{T})^2} \right) 
     \,  \nonumber\\ 
        &&{} 
        - \frac{16\,(S+\bar S)^2}{\vert Y^0\vert^4 [(T + {\bar
          T})^2 ]^2} \Big((\partial_{S} \Omega)^2
           \bar D_{\bar S} \partial_{\bar S} \Omega 
           +(\partial_{\bar S} \Omega )^2  D_{
          S}\partial_{S} \Omega \Big) \;.\nonumber\\&~& 
\end{eqnarray}
This expression is manifestly invariant under both S- and T-duality.
Subsequently we evaluate (\ref{eq:expand-tildeS-Omega}), and obtain
the following result, after again suppressing the tildes on the
right-hand side,
\begin{eqnarray}
  \label{eq:expansion-Omega}
  \Omega (S , {\bar S}) &\approx&{} \Omega ({S}, {\bar {S}})
  \nonumber\\ 
  &&{}
  + \frac{4 \, ({S} + {\bar{S}}) \,\big\vert\partial_S
    \Omega \big\vert^2} {|{Y}^0|^2 ({T} + {\bar {T}})^2} 
  \left( 1
      +  \frac{3\, ({S}+ {\bar {S}})}{|{Y}^0|^2 ({T} + {\bar {T}})^2}
    \, \partial_S \partial_{\bar S} \Omega  \right) \nonumber\\ 
  &&{}  +  \frac{6\, ({S} + {\bar {S}})^2 }{
      |{Y}^0|^4 [({T} + {\bar {T}})^2]^2} \, \Big
    ((\partial_{\bar S} \Omega )^2 \, D_S \partial_S \Omega  +
    (\partial_S \Omega  )^2 \, {\bar D}_{\bar S} \partial_{\bar S}
    \Omega   \Big) \;. 
\end{eqnarray}
This is manifestly invariant under S- and T-duality. We observe that
both \eqref{eq:hesse-1-ext} and \eqref{eq:expansion-Omega} depend
non-holomorphically on $\tilde Y^0$ (we reinstate the tilde to
indicate that we are discussing the new variables). The only exception
is the first term in \eqref{eq:expansion-Omega}, which is equal to
$\Omega$ and does not depend on $\tilde Y^0$, nor on its complex
conjugate.  All the terms arising in higher orders will always depend
on $\vert\tilde Y^0\vert$, and not on $\tilde Y^0$ or $\bar{\tilde
  Y}^0$, separately. This is because invariance under T-duality
dictates that each power of $(\tilde T + \bar{\tilde T})^2$ (arising
by power expanding in ${\cal V}$) has to appear multiplied by $|\tilde
Y^0|^2$. We return to the significance of this observation in later
sections.

\section{The Hesse potential at second order}
\label{sec:hesse-potential-2-order}
\setcounter{equation}{0}
In this section, we return to the general case based on \eqref{eq:F_I}
and we consider the Hesse potential, using the representation
\eqref{eq:GenHesseP2}. It consists of two parts which both transform
as proper functions under electric/magnetic duality. The first term is
equal to
\begin{eqnarray}
 \label{eq:hesse-1}
   -\mathrm{i} (\bar Y^I F_I-Y^I\bar F_I)&=&
   -\vert Y^0\vert^2 (S+\bar S)(T+\bar T)^2 +
   2\left(Y^0\,\frac{\partial\Omega}{\partial Y^0}+ \bar
   Y^0\,\frac{\partial\Omega}{\partial\bar Y^0} \right)\nonumber\\[.2ex]
   &&{}
   + 2\,(\bar Y^0\,\mathcal{U} + Y^0\,\bar{\mathcal{U}}) -
   2\,(Y^0+\bar Y^0) \,(T+\bar T)^a\,\mathcal{T}_a \;,
\end{eqnarray}
where we made use of \eqref{eq:F_I}. The second term contributing to
the Hesse potential is equal to
$4\,(\Upsilon\partial_\Upsilon\Omega+\bar
\Upsilon\partial_{\bar\Upsilon}\Omega)$. This term is separately
invariant under the dualities. Combining both terms and making use of
the homogeneity of the function $\Omega$, i.e., $2 \Omega = 2 \Upsilon
\, \Omega_{\Upsilon} + 2 {\bar \Upsilon}\, \Omega_{\bar \Upsilon} +Y^I
\, \Omega_I + {\bar Y}^I \, \Omega_{\bar I}$, it follows that the
Hesse potential takes the form,
\begin{eqnarray}
 \label{eq:Hesse-homog}
   \mathcal{H}(\tilde Y,\bar{\tilde Y},\Upsilon,\bar\Upsilon)
       &=&
   -\vert Y^0\vert^2 (S+\bar S)(T+\bar T)^2 +
   4\, \Omega(Y^0,\bar Y^0, S,\bar S, T,\bar T,\Upsilon,\bar
   \Upsilon)        \nonumber\\[.2ex]
   &&{}
   + 2\,(\bar Y^0\,\mathcal{U} + Y^0\,\bar{\mathcal{U}}) -
   2\,(Y^0+\bar Y^0) \,(T+\bar T)^a\,\mathcal{T}_a \;.
\end{eqnarray}
This result is written as a function of the old fields $Y^0$, $S$, and
$T^a$, which can be expressed in terms of the new fields by using
\eqref{eq:new-variables}. Expressing the first term in the new fields
$\tilde Y^0$, $\tilde S$ and $\tilde T^a$, generates contributions up
to fifth order in $\Delta Y^0$, $\Delta S$ and $\Delta T^a$, which,
upon iteration, can be expressed as a power series in $\mathcal{U}$
and $\mathcal{T}_a$. Here we will consider terms of first- and
second-order. As it turns out, the first-order term cancel against
those explicitly given in \eqref{eq:hesse-1}. This is a general
phenomenon which applies also to other models than the ones based on
\eqref{eq:F-0}. Making use of \eqref{eq:Delta-S-second-order},
\eqref{eq:Delta-T-second-order}, and \eqref{eq:Delta-Y0-second-order},
we obtain,
\begin{eqnarray}
 \label{eq:Hesse-approx}
   \mathcal{H}(\tilde Y,\bar{\tilde Y},\Upsilon,\bar\Upsilon)
   &\approx& 
   -\vert \tilde Y^0\vert^2 (\tilde S+\bar{\tilde S})(\tilde
   T+\bar{\tilde T})^2 +
   4\, \Omega(Y^0,\bar Y^0, S,\bar S, T,\bar T,\Upsilon,\bar
   \Upsilon) \nonumber\\[,2ex]
   &&{} + \frac{2\, \mathcal{T}_a\eta^{ab}\mathcal{T}_b}{S+\bar S}
   -\frac{8\,\vert\mathcal{U}\vert^2}{(S+\bar S)(T+\bar T)^2}
   +\cdots\,,
\end{eqnarray}
where the ellipses denote terms of third and higher order in
derivatives of $\Omega$. In this approximation the result is invariant
under S- and T-duality, as it should. The first term is manifestly
invariant. The invariance of the remaining three terms follows
directly from application of the relevant equations in
\eqref{eq:S-invariance}, \eqref{eq:T-invariance}, \eqref{eq:T-U-Sdual}
and \eqref{eq:T-U-Tdual}.

Subsequently we express $\Omega$ in terms of the new variables. This
can be done by Taylor expanding, using \eqref{eq:new-variables},
\begin{eqnarray}
 \label{eq:Taylor}
 &&\Omega(Y^0,\bar Y^0, S,\bar S, T,\bar T) \approx \Omega(\tilde
   Y^0,\bar{\tilde Y}^0, \tilde S,\bar{\tilde S}, \tilde
   T,\bar{\tilde T})
    - \Delta Y^0
   \left(\frac{\partial\Omega}{\partial
   Y^0}-\frac{\partial\Omega}{\partial \bar Y^0}\right)\Big\vert_\ast
   \nonumber \\[.2ex]
   &&{}\qquad
   -\Delta S \,\frac{\partial\Omega}{\partial S}\Big\vert_\ast
   -\Delta \bar S\, \frac{\partial\Omega}{\partial\bar S}\Big\vert_\ast
   - \Delta T^a \,\frac{\partial\Omega}{\partial T^a}\Big\vert_\ast
   - \Delta \bar T^a\, \frac{\partial\Omega}{\partial \bar T^a}
   \Big\vert_\ast + \cdots\;,
\end{eqnarray}
where we have suppressed the variables $\Upsilon$. The notation
$\vert_\ast$ indicates that the derivatives are taken at $\tilde Y^0$,
$\tilde S$ and $\tilde T$. However, in the approximation that we
adopted, this aspect is only relevant in higher orders. The terms
generated by the Taylor expansion turn out to be proportional to
$\vert\mathcal{U}\vert^2$ and $\mathcal{T}_a\eta^{ab} \mathcal{T}_b$.
Substituting the above result into \eqref{eq:Hesse-approx}, we obtain
\begin{eqnarray}
 \label{eq:Hesse-approx-2}
   \mathcal{H}(\tilde Y,\bar{\tilde Y},\Upsilon,\bar\Upsilon)
   &\approx& 
   -\vert \tilde Y^0\vert^2 (\tilde S+\bar{\tilde S})(\tilde
   T+\bar{\tilde T})^2 +
   4\, \Omega(\tilde Y^0,\bar{\tilde Y}^0, \tilde S,\bar{\tilde S},
   \tilde T,\bar{\tilde T}) \nonumber\\[.2ex]
   &&{} - \frac{2\, \mathcal{T}_a\eta^{ab}\mathcal{T}_b}{S+\bar S}
   + \frac{8\,\vert\mathcal{U}\vert^2}{(S+\bar S)(T+\bar T)^2} \,,
\end{eqnarray}
which holds to second order in $\Omega$ and derivatives thereof. 

In order to compare with other results we write $\Omega$ as a power
series in the Weyl background $\Upsilon$ (just as in
\eqref{eq:top-string}) and/or $\bar\Upsilon$, so that we may write,
\begin{equation}
 \label{eq:Omega-exp}
 \Omega(Y^0,\bar Y^0,S,\bar S,T,\bar T,\Upsilon,\bar \Upsilon) =
 \sum_{g=1}^\infty
 \;\Omega^{(g)}(Y^0,\bar Y^0,S,\bar S,T,\bar T) \,,
\end{equation}
where $\Omega^{(g)}$ is real and decomposable in monomials of the form
$\Upsilon^n\,\bar \Upsilon^{g-n}$, with $0\leq n\leq g$. For
conciseness, we will refrain from explicitly indicating the dependence
of the functions $\Omega^{(g)}$ on $\Upsilon$ and $\bar\Upsilon$. In
the approximation that we retain terms of second order of
$\Upsilon,\bar\Upsilon$, we will only have contributions from
$\Omega^{(1)}$ and $\Omega^{(2)}$ which we can regard as functions of
the new fields $\tilde Y^I$. Since the result depends then only on the
new fields $\tilde Y^I$, we can now consistently drop the distinction
between the variables $\tilde Y^I$ and $Y^I$ to simplify our
notation. Therefore, we suppress the tilde on the right-hand side in
the formula below,
\begin{eqnarray}
 \label{eq:Hesse-approx-3}
   \mathcal{H}(\tilde Y,\bar{\tilde Y},\Upsilon,\bar\Upsilon) &\approx&
   -\vert Y^0\vert^2 ( S+\bar{S})(
   T+\bar{T})^2 +
   4\, \Omega^{(1)}(S,\bar{ S}, T,\bar{T})
   \nonumber\\[.2ex]
   &&{}
   +4\, \Omega^{(2)}(Y^0,\bar{ Y}^0, S,\bar{S}, T,\bar{T})
   \nonumber\\[.2ex]
   &&{}
   -\left\{\frac2{(Y^0)^2}\left(\frac{\partial \Omega^{(1)}}{\partial
   T^a}
   \left[\frac1{S+\bar S}\, \frac{\partial \Omega^{(1)}}{\partial T_a}
   + \frac{4\,(T+\bar T)_a}{(T+\bar T)^2} \,\frac{\partial
   \Omega^{(1)}}{\partial S}\right]  \,\right) + \mathrm{h.c.}
   \right\} \nonumber\\[.2ex]
   &&{}
   + \frac{4\,(T+\bar T)^a (T+\bar T)^b}{|Y^0|^2 \, (S + \bar S)
   (T+\bar T)^2 }\,
 \left(2 \,\frac{\partial \Omega^{(1)}}{\partial T^a} \,
   \frac{\partial \Omega^{(1)}}{\partial{\bar T}^b}
   - \eta_{ab}\, \frac{\partial \Omega^{(1)}}{\partial T_c} \,
   \frac{\partial \Omega^{(1)}}{\partial {\bar T}^c} \right)
 \nonumber \\[.2ex]
   &&{}
   + \frac{ 8\,(S + \bar S)}{|Y^0|^2 \,(T + \bar T)^2 }\,
   \frac{\partial \Omega^{(1)}}{\partial S}
   \, \frac{\partial\Omega^{(1)}}{\partial \bar S} \;.
\end{eqnarray}
Note
that $\Omega^{(1)}$ does not depend on $Y^0$ and $\bar Y^0$ because of
homogeneity. The above result represents the Hesse potential to second
order in $\Upsilon,\bar\Upsilon$ and is the basis for our discussion
in the next section. Higher-order terms have a similar characteristics
as the terms we derived in section \ref{sec:heterotic-example}. They
always consist of the product of a number of first-order derivatives
of $\Omega$, times a number of higher-order derivatives of $\Omega$.

Let us comment on the various terms in \eqref{eq:Hesse-approx-3} (we
again reinstate the tildes to stress that we are dealing with the new
variables).  First we note that $\Omega^{(1)}$ and $\Omega^{(2)}$
define the higher-derivative corrections to the effective action, but
their arguments are not the moduli defined in that perspective, as
they are based on the Legendre transform. By construction the
expression should be duality invariant where the transformation rules
are the ones that pertain to the classical action (i.e. without
higher-derivative couplings), specified in \eqref{eq:ST-Sdual} and
\eqref{eq:T-S-Sdual}.

The first $\Omega$-independent term in \eqref{eq:Hesse-approx-3} is
duality invariant and so is $\Omega^{(1)}$ (provided we take
$\Upsilon$ real). This result applies to all orders, because of the
new variables that have been employed. It is known that $\Omega^{(2)}$
is not duality invariant \cite{Cardoso:2008fr}, and neither are the
terms proportional to $(\tilde Y^0)^{-2}$ or $(\bar{\tilde
  Y}^0)^{-2}$. On the other hand the two terms proportional to
$\vert\tilde Y^0\vert^{-2}$ are both S- and T-duality invariant. The
relevance of this decomposition will be explained in the next section.

As we emphasized already in section \ref{sec:introduction}, we
concentrate on S- and T-duality here in view of the fact that we have
only explicit information about models with a high degree of
symmetry. The decomposition in terms of $(\tilde Y^0)^{-2}$,
$(\bar{\tilde Y}^0)^{-2}$ and $\vert\tilde Y^0\vert^{-2}$ remains
relevant in the more general case, as $\tilde Y^0$ is the inverse
holomorphic coupling constant of the topological string. Consequently,
the terms proportional to $\vert\tilde Y^0\vert^{-2}$ cannot be part
of the twisted partition function of genus $g=2$, irrespective of
whether the model has certain duality invariances.

\section{The Hesse potential for specific models}
\label{sec:hesse-specific}
\setcounter{equation}{0}
In this section we consider the consequences of the results of the
previous sections in the context of a few specific models. As was
already mentioned in section \ref{sec:introduction}, there are only a
few models for which explicit results have been obtained for the
effective action and/or the topological string. The models that we
discuss are models with $N=4$ supersymmetry, cast in an $N=2$
description, and the FHSV model \cite{Ferrara:1995yx} (another
possible model is the STU model \cite{Sen:1995ff,Gregori:1999ns}, but
this is qualitatively similar to the FHSV model). As it turns out the
expression for $\Omega^{(1)}(S,\bar S, T,\bar T)$ coincides for both
the effective action, the Hesse potential and the topological string.

We begin with the $N=4$ supersymmetric model discussed in section
\ref{sec:heterotic-example} for which $\Omega = \Omega(S,\bar{ S})$
depends only on $S$ and $\bar S$. Particular examples are the
so-called CHL models \cite{Chaudhuri:1995fk}. Using
\eqref{eq:hesse-1-exact}, we obtain the following exact expression for
the Hesse potential,
\begin{eqnarray}
 \label{eq:Hesse-exact-N4}
   \mathcal{H}(\tilde Y,\bar{\tilde Y},\Upsilon,\bar\Upsilon) &=&
   - \vert{\tilde Y^0}\vert^2  ({\tilde S} + {\bar {\tilde S}})
   ({\tilde T}+ {\bar {\tilde T}})^2 \, \frac{1 + 2 \,({\cal V}
   + {\bar {\cal V}}) }{\sqrt{
   1 + 4 \,({\cal V}
   + {\bar {\cal V}}) + 4\, ({\cal V}
   - {\bar {\cal V}})^2 }   } \nonumber\\
   &&{} + 4 \, \Omega(S,\bar{ S}) \;.
\end{eqnarray}
Note that we have not constrained $\Omega(S,\bar S)$ other than that
it should be invariant under the S-duality group. The definition of
$\mathcal{V}$ is subtle and follows from (\ref{eq:cal-V-cont}) and
(\ref{eq:def-lambda}). The result was evaluated up to terms cubic in
$\Omega$-derivatives, and combining the explicit results
(\ref{eq:hesse-1-ext}) and (\ref{eq:expansion-Omega}), one obtains the
following expression (where, again, we suppressed the tildes on the
right-hand side of the equation),
\begin{eqnarray}
  \label{eq:Hesse-approx-N4}
  \mathcal{H}(\tilde Y,\bar{\tilde Y},\Upsilon,\bar\Upsilon)\!\!
  &\approx& \!\!{}
  -\vert Y^0\vert^2 (S+\bar{S})(T+\bar{ T})^2 +
  4\, \Omega(S,\bar{S})\nonumber\\[.4ex]
  &&{}
  +\frac{8 \, (S+\bar{S})\,
    \big\vert\partial_S\Omega\big\vert^2} {|{Y}^0|^2
    (T+\bar{T})^2} \left(1 +
    \frac{2\,(S+\bar{S})\, \partial_S \partial_{\bar S}\Omega}
    {\vert{Y}^0|^2 (T+\bar{T})^2} \right) \nonumber\\[.2ex]
        &&{} 
        + \frac{8\,(S+\bar S)^2}{\vert Y^0\vert^4 [(T + {\bar
          T})^2 ]^2} \Big((\partial_{S} \Omega)^2
           \bar D_{\bar S} \partial_{\bar S} \Omega 
           +(\partial_{\bar S} \Omega )^2  D_{
          S}\partial_{S} \Omega \Big) \;, 
\end{eqnarray}
to third order in $\Omega$ and derivatives thereof.  This expression
exhibits the dependence of the Hesse potential on both ${\tilde Y}^0$ and its
complex conjugate (here we again reinstated the tilde). As we already argued in section
\ref{sec:heterotic-example}, the higher-order terms will depend only
on $\vert {\tilde Y}^0\vert$, and no longer on ${\tilde Y}^0$ and $\bar {\tilde Y}^0$ separately.

For completeness we recall the expression for $\Omega_k$ for the CHL
models distinguished by an integer label $k$. As discussed in
\cite{Jatkar:2005bh} the function $\Omega_k$ can be expressed in terms
of the unique cusp forms of weight $k+2$ associated with the S-duality
group $\Gamma_1(\tilde N)\subset\mathrm{SL}(2;\mathbb{Z})$, defined by
$f^{(k)}(S) = \eta^{k+2}(S)\,\eta^{k+2}(\tilde N S)$ where,
\begin{equation}
  \label{eq:f-S-dual}
  f^{(k)} (S^\prime) = \Delta_{\mathrm{S}}^{\,k+2}  \, f^{(k)}(S)\,.
\end{equation}
The result for $\Omega_k$ then takes the following form
\cite{LopesCardoso:2006bg},
\begin{equation}
  \label{eq:Omega-het}
  \Omega_k(S,\bar S,\Upsilon,\bar\Upsilon) =
  {}\frac{1}{256\,\pi} \Big[\Upsilon \ln f^{(k)}(S)  +
\bar\Upsilon \ln f^{(k)}(\bar S) + \ft12(\Upsilon+\bar \Upsilon)
\ln (S+\bar S)^{k+2} \Big] \,.
\end{equation}
Note that this result agrees with the terms obtained for the
corresponding effective actions (see, for instance,
\cite{Harvey:1996ir,Gregori:1997hi}). These models are invariant under
the S-duality group $\Gamma_1(\tilde N)
\subset\mathrm{SL}(2;\mathbb{Z})$, which is generated by 
\eqref{eq:ST-S-full} with the transformation parameters restricted to
$c = 0\mod\tilde N$ and $a,d=1\mod{\tilde N}$.

Next, we consider the $N=2$ supersymmetric FHSV model. Its type-II
realization corresponds to the compactification on the Enriques
Calabi-Yau three-fold, which is described as an orbifold
$(\mathrm{T}^2\times\mathrm{K3})/\mathbb{Z}_2$, where $\mathbb{Z}_2$
is a freely acting involution. The massless sector of the
four-dimensional theory comprises 11 vector supermultiplets, 12
hypermultiplets and the $N=2$ graviton supermultiplet. The classical
moduli space of the the vector multiplet sector equals the
special-K\"ahler space,
\begin{equation}
 \label{eq:4vector-special-K}
 \mathcal{M}_{\mathrm{vector}}=
 \frac{\mathrm{SL}(2)}{\mathrm{SO}(2)}\times
 \frac{\mathrm{O}(10,2)}{\mathrm{O}(10)\times\mathrm{O}(2)}\,.
\end{equation}
Its two factors are associated with $\mathrm{T}^2/\mathbb{Z}_2$ and
the $\mathrm{K3}$ fiber, and the special coordinates for these two
spaces will be denoted by $S$ and $T^a$, respectively. 

At first order in the Weyl background, the solution to
\eqref{eq:S-invariance} and \eqref{eq:T-invariance} for the FHSV model
is known from threshold corrections and from the topological string
side \cite{Harvey:1996ts,Klemm:2005pd}.  It takes the form
\cite{Cardoso:2008fr},
\begin{eqnarray}
 \label{eq:Omega-FHSV}
 \Omega^{(1)}(S,\bar S,T, \bar T, \Upsilon,\bar\Upsilon)
 &=&{}
 \frac{1}{256\,\pi} \Big[\tfrac12 \Upsilon
 \ln[\eta^{24}(2S)\,\Phi(T)]
 +\tfrac12\bar\Upsilon \ln[\eta^{24}(2\bar S)\,\Phi(\bar T)]
 \nonumber\\
 &&{}\hspace{1cm}
 +(\Upsilon+\bar \Upsilon)
 \ln [(S+\bar S)^3 (T+\bar T)^a\eta_{ab}(T+\bar T)^b]\Big]
 \,.
\end{eqnarray}
For real values of $\Upsilon$, this result is invariant under
S-duality, which constitute the $\Gamma(2)$ subgroup of
$\mathrm{SL}(2;\mathbb{Z})$, defined by $a,d= 1\mod 2$ and $b,c =
0\mod 2$ in \eqref{eq:ST-Sdual}. The result is also invariant under
the T-duality group $\mathrm{O}(10,2;\mathbb{Z})$ in view of the fact
that $\Phi(T)$ is a holomorphic automorphic form of weight 4
\cite{Borcherds:1996}, transforming under the T-duality transformation
$T^a \rightarrow T^a\, [T^2]^{-1}$ as
\begin{equation}
 \label{eq:Phi-T}
 \Phi(T)\to [T^2]^{4} \,\Phi(T)\,.
\end{equation}
Clearly, \eqref{eq:Omega-FHSV} can be written as the sum of two
invariant functions, one of $S$ and $\bar S$ and one of $T^a$ and
$\bar T^a$, respectively, which both contain non-holomorphic terms
that are crucial for the duality invariance.  Observe that the duality
invariance of $\Omega^{(1)}$ is only realized for real values of
$\Upsilon$.  Therefore we do not know a priori whether to write
$\Upsilon$ or its complex conjugate. The way in which this potential
ambiguity has been resolved, is by assuming that purely holomorphic
terms are always accompanied by a power of $\Upsilon$ and purely
anti-holomorphic terms by a power of $\bar\Upsilon$, whereas for the
mixed terms we assign $\Upsilon$ and $\bar\Upsilon$ such as to
preserve the reality properties of $\Omega$ for complex $\Upsilon$.

At second order in the Weyl background, the solution to
\eqref{eq:S-invariance} and \eqref{eq:T-invariance} for the FHSV model
takes the following form \cite{Cardoso:2008fr}, up to an S- and
T-duality invariant function,
\begin{equation}
 \label{eq:Om2res}
 \Omega^{(2)} =
 - \frac{G_2 (2S)}{(Y^0)^2}\,
 \frac{\partial \Omega^{(1)}}{\partial T^a}\,
 \frac{\partial \Omega^{(1)}}{\partial T_a}
 -\frac{1}{4(Y^0)^2}\,
 \frac{\partial \ln\Phi(T)}{\partial T_a}\,
 \frac{\partial\Omega^{(1)}}{\partial T^a} \, \frac{\partial
 \Omega^{(1)}}{\partial S}   + {\rm c.c} \;,
\end{equation}
where $G_2 (2 S) = \tfrac12 \partial_S \ln \eta^2 (2S)$.  Observe that
$\Omega^{(2)}$ is not duality invariant. Inserting it into
\eqref{eq:Hesse-approx-3}, one obtains (the right-hand side is
expressed exclusively in terms of the new variables but we again
suppress the tilde on the right-hand side for clarity),
\begin{eqnarray}
 \label{eq:Hesse-approx-3-FHSV}
   \mathcal{H}(\tilde Y,\bar{\tilde Y},\Upsilon,\bar\Upsilon)
   &\approx& 
   -\vert Y^0\vert^2 ( S+\bar{S})(
   T+\bar{T})^2 +
   4\, \Omega^{(1)}(S,\bar{ S}, T,\bar{T})
   \nonumber\\[.2ex]
   &&{}
   - \left[
     \frac{4\, {\hat G}_2(2S, 2 \bar S)}{(Y^0)^2} \,
     \frac{\partial \Omega^{(1)}}{\partial T_a} \,
     \frac{\partial \Omega^{(1)}}{\partial T^a} \right. \nonumber\\
   &&{} \quad \quad \left.
     + \frac{1}{(Y^0)^2} \,
     \frac{\partial\log\left[\Phi(T)\, [(T+\bar T)^2]^4\right]}
   {\partial T_a} 
     \, \frac{\partial \Omega^{(1)}}{\partial T^a}
     \, \frac{\partial \Omega^{(1)}}{\partial S}
     + {\rm c.c.}  \right]  \nonumber\\ 
       &&{}
       + \frac{4\,(T+\bar T)^a (T+\bar T)^b}{|Y^0|^2 \, (S + \bar S)
         (T+\bar T)^2 }\,
       \left(2 \,\frac{\partial \Omega^{(1)}}{\partial T^a} \,
         \frac{\partial \Omega^{(1)}}{\partial{\bar T}^b}
         - \eta_{ab}\, \frac{\partial \Omega^{(1)}}{\partial T_c} \,
         \frac{\partial \Omega^{(1)}}{\partial {\bar T}^c} \right)
       \nonumber \\
       &&{}
       + \frac{ 8\,(S + \bar S)}{|Y^0|^2 \,(T + \bar T)^2 }\,
       \frac{\partial \Omega^{(1)}}{\partial S}
       \, \frac{\partial\Omega^{(1)}}{\partial \bar S} \;,
\end{eqnarray}
where $\hat G_2(2 S,2 \bar S)= G_2(2 S) + [2\,(S+\bar S)]^{-1}$. In
this expression, $\Omega^{(1)}$ is given by \eqref{eq:Omega-FHSV},
with the old variables replaced by the new ones. This result is
manifestly invariant under S- and T-duality, as it should.
Furthermore, the terms proportional to $(Y^0)^{-2}$ can be combined by
noting that, for real values of $\Upsilon$, we have the identities,
\begin{eqnarray}
  \label{eq:partial-Omega-G}
  \frac{\partial\Omega^{(1)}}{\partial S} &=& \frac{24\, \Upsilon}
  {512\,\pi} \,\hat G_2(2S,2\bar S) \,,
   \nonumber\\ 
   \frac{\partial\Omega^{(1)}}{\partial T^a} &=& \frac{\Upsilon}
  {512\,\pi} \,\frac{\partial\log\left[\Phi(T)\, [(T+\bar T)^2]^4\right]}
  {\partial T^a} \,.
\end{eqnarray}
Therefore these terms can be rewritten as (suppressing the tildes on both sides),
\begin{equation}
  \label{eq:Hesse-y02}
  \mathcal{H}(Y,\bar{Y},\Upsilon,\bar\Upsilon)\Big\vert_{(Y^0)^{-2}} = 
  \frac{28\,{\hat G}_2(2S, 2 \bar S)}{(Y^0)^2} \, 
  \frac{\partial \Omega^{(1)}}{\partial T_a} \,
  \frac{\partial \Omega^{(1)}}{\partial T^a}   \,. 
\end{equation}
This is consistent with the result found for the topological string
\cite{Grimm:2007tm}, apart from the overall normalization. However, we
can change the normalization by including the same duality invariant
expression into $\Omega^{(2)}$ with a different coefficient. Since
(\ref{eq:Om2res}) has only been determined up to a duality invariant
function, and since we have no independent knowledge of the invariant
parts in $\Omega^{(2)}$, the overall normalization given in
(\ref{eq:Hesse-y02}) is therefore ambiguous. At this point we should
recall that the topological string partition functions are derived
from integrating the holomorphic anomaly equations
\cite{Grimm:2007tm}, so that the results are in principle determined
up to holomorphic contributions. On the other hand, results such as
(\ref{eq:Om2res}) have been obtained from requiring covariance under
duality transformations, and therefore they determine the
$\Omega^{(g)}$ up to duality invariant terms. Usually the invariant
terms are non-holomorphic, so that combining the two methods could
potentially remove the ambiguities. However, there can also be
holomorphic, invariant functions, which would be missed in both
approaches. As in \cite{Grimm:2007tm}, one may be able to remove some
of these ambiguities by making use of knowledge of the boundary
behaviour or certain asymptotic conditions, but at present this is not
really known.

Based on the previous arguments, and modulo the ambiguity noted above
in the normalization of the genus-2 contribution, the topological
string partition function would correspond to the following function
$F^\mathrm{top}$, which can be viewed as the analogue of
(\ref{eq:F-decomposition}) for the effective action,
\begin{eqnarray}
  \label{eq:top-parti-FHSV}
  &&F^\mathrm{top} \approx {}\mathrm{i} (Y^0)^2 S\, T^2 -
  \frac{\mathrm{i}}{2\, \pi}  
  \ln\left[\vert\eta^{24}(2S)\vert\,(S+\bar S)^6 \,\vert\Phi(T)\vert\,
    [(T+\bar T)^2]^2  \right] \nonumber\\[.2ex] 
  &&{}   
  + a\;\frac{\mathrm{i}\, {\hat G}_2(2S, 2 \bar S)}{(2\,\pi\,Y^0)^2} \, 
  \frac{\partial \ln\left[\vert\Phi(T)\vert\,[(T+\bar T)^2]^2\right]}
  {\partial T_a} \, 
  \frac{\partial \ln\left[\vert\Phi(T)\vert\,[(T+\bar T)^2]^2\right]} 
  {\partial T^a}   \,, 
\end{eqnarray}
where we set $\Upsilon=-64$. Here the real constant $a$ represents the
ambiguity discussed above, which can be fixed by imposing the
holomorphic anomaly equation. We stress that the arguments used in
(\ref{eq:top-parti-FHSV}) refer to the new variables $\tilde Y^0$,
$\tilde S$ and $\tilde T^a$, and that the topological string coupling
constant $g_\mathrm{s}$ is inversely proportional to $\tilde Y^0$.

One may now wonder what the role is of the contributions to the Hesse
potential that do not depend holomorphically on $\tilde Y^0$. These
are the terms in (\ref{eq:Hesse-approx-3-FHSV}) proportional to
$\vert\tilde Y^0\vert^{-2}$, which are not part of the topological
string partition function, as they do not depend holomorphically on
the topological string coupling. Here we reinstated the tilde to
emphasize that we are dealing with the new variables. Obviously these
terms are duality invariant, and therefore their normalization factors
can in principle be changed upon including similar invariant terms
into $\Omega^{(2)}$. Hence the normalization remains ambiguous, which
makes it hard to assess the relevance of these terms.

At this point let us return to the $N=4$ supersymmetric models and
consider the expression for the Hesse potential
(\ref{eq:Hesse-approx-N4}). In this case there is only a dependence on
$\vert \tilde Y^0\vert$ and the only contribution to the topological
string originates from $\Omega(\tilde S,\bar{\tilde S})$. This result
is in agreement with known results for the topological string
partition function \cite{Bershadsky:1993cx}. The normalization of this
genus-1 term is unambiguous, which lends support to the discussion of
the FHSV model given above. Concerning the terms that depend on
negative powers of $\vert \tilde Y^0\vert$, in principle such duality
invariant terms can also be present in $\Omega$ as contributions to
the effective action, and their presence would affect the
normalization factors of the corresponding terms in the Hesse
potential. However, in that case the result would no longer be
consistent with the initial assumption that was made in section
\ref{sec:heterotic-example}, namely that we assumed from the beginning
that $\Omega$ depends only on $S$ and $\bar S$, thus excluding any
other additional terms in the effective action. This initial
assumption was partly a matter of convenience, and it is difficult to
fully exclude other starting points at this stage.

In closing we conclude that, generically, the Hesse potential may
contain terms that are non-holomorphic in the topological string
coupling constant, and that these do not exclusively originate from
the non-holomorphic corrections in the effective action.  However, due
to lack of data on both the effective action and on the topological
string side, we cannot at present draw a definite conclusion about the
presence of such terms in the Hesse potential. Our findings do not, at
this stage, contradict the idea that the Hesse potential could
actually coincide with the topological string. Should this be the
case, this will have calculable implications for the effective action,
which in principle can be worked out explicitly by means of the
iterative method proposed in this paper.

\subsection*{Acknowledgements}
We acknowledge helpful discussions with Albrecht Klemm, Thomas Grimm,
Thomas Mohaupt, Ashoke Sen and Marcel Vonk. The work of G.L.C. is
supported by the Funda\c{c}\~{a}o para a Ci\^{e}ncia e a Tecnologia
(FCT/Portugal) and by a partnership grant PHYS0167 of the Alexander
von Humboldt Stiftung.

\providecommand{\href}[2]{#2}
\begingroup\raggedright\endgroup
\end{document}